\documentclass[twocolumn]{autart}    

\usepackage{hyperref}
\usepackage{cite}

\usepackage{graphicx} 
\usepackage{epsfig} 
\usepackage{mathptmx} 
\usepackage{algorithm}
\usepackage{algpseudocode}
\usepackage{times} 
\usepackage{amsmath} 
{
  \theoremstyle{plain}
  \newtheorem{assumption}{Assumption}
}
\usepackage{amssymb}  
\usepackage{xcolor}
\usepackage{subfigure}
\usepackage{lipsum}
\usepackage{balance}
\usepackage{flushend}
\usepackage{array}

\newtheorem{theorem}{Theorem}
\newtheorem{lemma}{Lemma}
\newtheorem{problem}{Problem}
\newtheorem{remark}{Remark}
\newtheorem{definition}{Definition}

\DeclareMathAlphabet{\pazocal}{OMS}{zplm}{m}{n}
\newcommand{\norm}[1]{\left\lVert#1\right\rVert}
\newcommand{\qApprox}{\textrm{quad}}

\newcommand{\R}{\mathbb{R}}
\newcommand{\J}{\pazocal{J}}

\newcommand{\sigmoid}{\text{sigmoid}}

\allowdisplaybreaks

\begin{document}

\begin{frontmatter}

\title{Newton's Method and Differential Dynamic Programming for Unconstrained Nonlinear Dynamic Games}


\author[Minneapolis]{Bolei Di}\ead{dixxx047@umn.edu},    
\author[Minneapolis]{Andrew Lamperski}\ead{alampers@umn.edu},               

\address[Minneapolis]{Electrical Engineering Department, 200 Union St SE, Minneapolis, MN 55455, Minneapolis}  

%
%
%
%
%

\begin{abstract}
  Dynamic games arise when multiple agents with differing objectives
  control a dynamic system.
  They model a wide variety of applications in economics,
  defense, energy systems and etc.
  However, compared to single-agent control problems, the
  computational methods for dynamic games are relatively limited.
  As in the single-agent case, only specific dynamic games can
  be solved exactly, so approximation algorithms are
  required.
  In this paper, we show how to extend a recursive Newton's algorithm and the popular differential dynamic programming (DDP) for single-agent optimal control to the case of full-information non-zero sum dynamic games.
  In the single-agent case, the convergence of DDP is proved by comparison with
  Newton's method, which converges locally at a quadratic rate.
  We show that the iterates of Newton's method and DDP are
  sufficiently close for the DDP to inherit the quadratic convergence
  rate of Newton's method.
  We also prove both methods result in an open-loop Nash equilibrium and a local feedback $O(\epsilon^2)$-Nash equilibrium.
  Numerical examples are provided.
\end{abstract}

\end{frontmatter}


\section{Introduction}
We study finite-horizon, unconstrained, discrete-time dynamic games or multistage games \cite{krawczyk2018multistage} with full information in this paper. Such games arise when multiple agents with differing objectives act upon one same dynamic system. The system can be naturally discrete-time or emerge from discretization of a differential game \cite{basar2018a, sethi2019differential, bauso2016game, bressan2011noncooperative}. 
Dynamic games have many applications including pursuit-evasion
\cite{rusnak2005lady}, active-defense
\cite{prokopov2013linear,garcia2014cooperative}, economics
\cite{el2013dynamic} and the smart grid \cite{zhu2012differential}.
Despite a wide array of applications, the computational methods for
dynamic games are considerably less developed than the single-agent
case of optimal control.

\subsection{Methods Overview}
\label{sec:overview}
Dynamic games extend optimal control to multiple agents aiming at
optimizing different objective functions. The  most common solution
concepts are \textit{open-loop Nash equilibria (OLNE)} and
\textit{feedback Nash equilibria (FNE)}  \cite{basar1999dynamic,
basar2018a, krawczyk2018multistage}. Early works on OLNEs gave conditions for existence and uniqueness of OLNEs for convex cost games \cite{rosen1965existence}\cite{friedman1979entry}\cite{fershtman1984capital}.
Most pioneer works suggested using direct gradient
descent method to solve for the equilibria. Our work is along the
approach of dynamic programming, Bellman recursion and
quadratic approximation. In particular, we extend the classic Newton method and
the differential dynamic programming (DDP) method.

Extending both the classic Newton method and DDP to dynamic problems
and their analysis is of great theoretical and practical
interest. The proposed stagewise Newton method first approximates
the original problem with a local quadratic dynamic game, then performs
a Bellman recursion of the approximated game, while the DDP method
solves the quadratically approximated Bellman recursion for the
original game. Both methods find an OLNE and approximate local FNE
\cite{haurie2012games, krawczyk2018multistage}. Other than deriving
their basic algorithmic forms, we prove that both the algorithms
converge quadratically in the neighborhood of strict stationary points
and provide sufficient conditions for the stationary points to be
Nash equilibria. While the convergence of Newton's method is locally
quadratic and well-established, to prove convergence of DDP method, we
extend arguments from \cite{murray1984differential,
  dunn1989efficient}, which relate DDP iterates to those of Newton's
method, to the case of dynamic games.

%

\subsection{Our Contribution}
We extend the numerical methods for dynamic games by offering
dedicated, practical algorithms for solving locally approximated
open-loop and feedback Nash equilibria for unconstrained nonlinear
systems, extending the classic Newton method and differential
dynamic programming. We prove that the algorithms inherit the
quadratic convergence rate of Newton's method
\cite{nocedal2006numerical} and provided sufficient conditions for the
solutions to be an OLNE and a locally approximated FNE. Compared to
our methods, other dedicated numerical methods for dynamic games
suffer from restricted scope or limited development, and static
methods suffer from high computational complexity.

\subsection{Paper Outline}
A literature survey is offered in Section~\ref{sec:lit_survey}. The general problem is formulated in Section~\ref{sec:problem}.
The  algorithms are described in Sections~\ref{sec:alg_nm} and~\ref{sec:alg_ddp}.  Convergence proof is sketched and equilibria are studied in Section~\ref{sec:convergence}. Section~\ref{sec:implementation_details} discusses implementation details.
Numerical examples are described in Section~\ref{sec:example}.
Conclusions and future directions are discussed in Section~\ref{sec:conclusion} while the detailed proofs are given in the appendix.

\section{Literature Survey}
\label{sec:lit_survey}
This section gives an overview of related numerical methods for dynamic
games. %
We will discuss methods for static games, including general Nash equilibrium
problems (GNEP) \cite{facchinei2007generalized_survey},
Newton's method \cite{facchinei2009generalized}, Nikaido-Isoda relaxation
algorithms \cite{berridge1997relaxation, krawczyk2000relaxation,
  krawczyk2007numerical, contreras2004numerical, krawczyk2005coupled},
and extremum seeking
\cite{frihauf2012nash,frihauf2013finite,frihauf2013nash}.
We will also discuss methods for special dynamic games, such as
linear-quadratic games
\cite{bressan2018stability, engwerda2017numerical,
  kebriaei2018discrete, engwerda2012feedback, krikelis1971solution,
  duncan2015some}, potential games
\cite{gonzalez2016survey,gonzalez2013discrete,gonzalez2014dynamic,mazalov2017linear,zazo2015new,zazo2015dynamic}, and
zero-sum games \cite{sun2015game, sun2016stochastic}. 
Finally,  we will discuss general methods based on Pontryagin's Maximum Principle
\cite{krawczyk2018multistage, basar2018a,bauso2016game,
  cacace2011numerical}. 
These
existing methods for games suffer from different reasons when applied to nonlinear
dynamic games or only handle special cases.
For a broad overview of recent
developments, see \cite{basar2018handbook}.

%
%


General Nash equilibrium problems (GNEPs) are  games with
constraints that may be coupled \cite{facchinei2007generalized_survey}. GNEPs are
reformulated to a set of necessary conditions via KKT conditions,
which is in the form of variational inequalities (VI). These
inequalities can be
solved via generic VI methods or classic feasibility problem methods,
such as Newton's method \cite{facchinei2009generalized} or others
\cite{bigi2018nonlinear}. In particular, Newton's method converts the
complementarity conditions to equality constraints via complementarity
functions. While these static methods for GNEPs
 can be applied to dynamic games, the iterations will have
computational complexity of $O(T^3)$ where $T$ is the number of
stages, because of the unexploited dynamic structure.
Our proposed Newton's method is closely in-line with Newton's
method for quasi-variational inequalities \cite{facchinei2009generalized} but more specialized
and faster because they exploit the dynamic structure.

The Nikaido-Isoda relaxation algorithm (NIRA) is another
method for solving GNEPs \cite{berridge1997relaxation,
  krawczyk2000relaxation, krawczyk2007numerical,
  contreras2004numerical, krawczyk2005coupled}. The iteration of this
method is based on weighted average of the current action and the
\textit{best response function}, which returns the set of players'
actions that minimize each of their cost unilaterally given the
current actions. The method converts the relatively hard root-finding
nature of solving for a NE to an optimization problem. However, the
convergence conditions are very restrictive.
It also does not utilize the dynamic structure,
therefore does not scale well w.r.t. number of stages when applied to
dynamic games.

Methods for finding Nash equilibria of static games via
extremum seeking were presented in
\cite{pan2004sliding, frihauf2012nash,frihauf2013finite,frihauf2013nash}. In
particular, the controllers drive the system to a Nash
equilibrium. The work expands from linear system to general nonlinear
systems. For these works, each agent only requires measurements of its
own
cost. 
Our method requires each agent to have explicit model
information, but gives equilibria for finite-horizon dynamic games. This is
particularly important for games in which trajectories from initial to
final states are desired.

As in optimal control, linear-quadratic (LQ) systems for games are
well-understood compared to general systems and serve as the backbone for
many solution methods \cite{bressan2018stability,
  engwerda2017numerical, kebriaei2018discrete, engwerda2012feedback,
  krikelis1971solution, duncan2015some}. The existence of FNEs for
linear-quadratic systems, and their analytic computation by
 coupled Riccati equations, is well understood
\cite{basar1999dynamic, basar1976uniqueness,
  engwerda2017numerical}.
The solution has also been extended to infinite
horizon and distributed information cases \cite{engwerda2000feedback,
  lin2013differential}. For a detailed description of the method for
solving linear-quadratic games see \cite{haurie2012games}; for the
complete set of sufficient conditions for discrete-time
linear-quadratic games see \cite{basar1999dynamic}.

In a potential game, a single potential function can be used to
describe the marginal costs for each player
\cite{dechert1997non,gonzalez2016survey,gonzalez2013discrete,gonzalez2014dynamic,mazalov2017linear,zazo2015new,zazo2015dynamic}. Based
on this property, potential games can often be solved using methods of
single-agent optimization or optimal control.
This line of work has been extended to constrained
stochastic dynamic potential games \cite{zazo2016dynamic}. However, the prerequisite that the game problem has a potential function is very restrictive.
Zero-sum game is another class of well-studied problems. Two player zero-sum differential games date back to the work of Issacs \cite{isaacs1999differential}.
Extensions such as stochastic zero-sum dynamic games also exist
\cite{exarchos2018stochastic, kushner2002numerical}.
This stream of work is very closely related to robust
control, in which a controller aims to perform well in the worst-case
\cite{diehl2004robust}. Work closely related to this paper is \cite{sun2015game,
  sun2016stochastic}, which applies DDP to zero-sum games. Our paper can be seen as a generalization
of \cite{sun2016stochastic} to multi-player nonzero-sum games with
theoretical justification.


The standard solution method for an OLNE is via Pontryagin's Minimum
Principle (PMP) for either continuous or discrete-time problems, as
recognized by the community \cite{bauso2016game, basar2018a,
  krawczyk2018multistage, cacace2011numerical, sethi2019differential,
  bressan2011noncooperative, carlson2016infinite}. Although the PMP
allows us to analyze the existence of solution and solve for
analytical solutions for a few simple games, the resulting boundary
value problem (BVP) with optimization is, in general, hard to solve
\cite{krawczyk2018multistage}. A more approachable reformulation of
the necessary conditions is concatenating the KKT conditions of each
player \cite{facchinei2007generalized_survey, krawczyk2018multistage},
in which case, we arrive at a structured nonlinear programming (NLP),
or \textit{feasibility problem}. Though it has been known for years
that such necessary conditions exist for games, we have not found
works on developing specialized algorithms for solving these
conditions, and generic solvers suffer from high complexity since they do  not
utilize the dynamic structure. Unlike its counterpart in optimal
control, the KKT conditions for games require users to solve a
root-finding problem, for which the conditions for existence of
solution and conditions for convergence of algorithm have not been
developed.


\section{Deterministic Nonlinear Dynamic Game Problem Formulation}
\label{sec:problem}
In this section, we introduce deterministic finite-horizon nonlinear game problem, the notations for the paper, the dynamic programming solution and convergence criterion of our proposed method.

\subsection{Problem Formulation}
\begin{problem}
  \label{problem:original}
  {\it
    \textbf{Nonlinear dynamic game} \\
    Each player tries to minimize their own cost
    \begin{align}
     \label{eq:cost}
      J_{n,t} (x, u) = \sum_{k = t}^T c_{n, k}(x_k, u_{:,k}), \quad n = 1, 2, ..., N
    \end{align}
    Subject to dynamic constraints
    \begin{subequations}
      \label{eq:dynamics}
      \begin{align}
        & x_{k+1} = f_k(x_k, u_{:,k}), \quad k = t, t + 1, \cdots, T-1 \\
        & x_0 \textrm{ is fixed.}
      \end{align}
    \end{subequations}
  }
\end{problem}
Here, $0 \leq t \leq T$ is the starting point for a game.
When  $t = 0$, we call it the \textit{full game}, and $t > 0$, a \textit{subgame $t$}. As indicated by the notations, we consider a full game of $T + 1$ steps played by $N$ players. The state of the system at time $k$ is denoted by $x_k\in
\mathbb{R}^{n_x}$.
Player $n$'s
input at time $k$ is given by $u_{n, k} \in
\mathbb{R}^{n_{u_{n}}}$. The vector of all players' actions at time $k$ is
denoted by $u_{:, k} = [u_{1,k}^\top, u_{2,k}^\top, \ldots,
u_{N,k}^\top]^\top \in \mathbb{R}^{n_u}$.
The cost for player $n$ at time $k$ is $c_{n,k}(x_k,u_{:,k})$.
In later analysis, some other notations will be helpful.
The vector player $n$'s
actions over all time is denoted by $u_{n,:} = [u_{n,0}^\top, u_{n,1}^\top, \ldots,
u_{n,T}^\top]^\top$. The vector of all actions other than those of
player $n$ is denoted by
$u_{-n,:} = [u_{1,:}^\top,\ldots,u_{n-1,:}^\top,u_{n+1,:}^\top,\ldots,u_{N,:}^\top]^\top$.
The vector of all states is denoted by
$x=[x_0^\top,x_1^\top,\ldots,x_T^\top]^\top$ while the vector of all inputs
is given by $u = [u_{1,:}^\top, u_{2,:}^\top, \ldots,
u_{N,:}^\top]^\top$.

Note that since the dynamics are
deterministic, the cost for each player can be expressed as a function
of all actions and the initial state, i.e. $J_{n, t}( x_t, u_{:,t:})$. Note that the dynamics are implicitly substituted to eliminate the dependency on $x$ when we use $J_{n, t}(x_t, u_{:,t:})$ and the subscript $t$ is omitted when we refer to the values of the full game. We assume $J_{n, t}(x_t, u_{:,t:})$ is twice differentiable. One set of sufficient conditions for the differentiablity of $J_{n, t}(x_t, u_{:,t:})$ is that both the cost $c_{n,k}(x_k, u_{:,k})$ and the dynamics $f_k(x_k, u_{:,k})$ share at least the same differentiability, which is not very restrictive since most physical systems are governed by ordinary differential equations.


\subsection{Local Open-loop Nash Equilibrium}
\label{sec:olne}

When discussing open-loop equilibria, we will fix the initial
condition, $x_0$ and the initial time $t=0$. For more compact
notation, we will drop the dependence on $t$ and $x_0$.

\begin{definition} \label{def:olne}
{\it
\textbf{(Local) open-loop Nash equilibrium} \\
  A \emph{local Nash equilibrium} (OLNE) for the full game problem
  \ref{problem:original} is a set of inputs $u^{\star}$ such that
  \begin{equation}
    \label{eq:olne}
    J_{n}( u_{n,:}, u_{-n, :}^{\star}) \geq J_{n}(u^{\star}), \  n = 1, 2, \ldots, N
  \end{equation}
  for all $u_{n,:}$. Furthermore, if \eqref{eq:olne} only holds for $u_{n, :}$ in a neighborhood of $u_{n,:}^\star$, it is called a \emph{local open-loop equilibrium}.
}
\end{definition}

The equilibrium is called a \emph{strict local Nash equilibrium} if
the inequality in \eqref{eq:olne} is strict for all $u_{n,:}\ne
u_{n,:}^\star$ in a neighborhood of $u_{n,:}^\star$. For unconstrained
games, the following problem gives necessary conditions for a local
Nash equilibrium:

\begin{problem}
  {\it
    \textbf{Necessary conditions.} Find $u^\star$ such that
    \begin{align}
      \label{eq:necessary}
      \pazocal{J}(u^\star) \equiv \left.
      \begin{bmatrix}
        \frac{\partial J_{1}}{\partial u_{1,:}} & \frac{\partial J_{2}}{\partial u_{2,:}} & \cdots & \frac{\partial J_{N}}{\partial u_{N,:}}
      \end{bmatrix}^\top\right|_{u^\star} = 0
    \end{align}
  }
\end{problem}

A trajectory of actions $u$ is referred to as a \textit{stationary point} satisfying \eqref{eq:necessary}. Solving such necessary conditions is standard which also arises in other works \cite{facchinei2007generalized, dutang2013survey, krawczyk2018multistage, basar2018handbook}.

\subsection{Feedback Nash Equilibrium}
\label{sec:fne}
In the case when state feedback information is available, feedback Nash equilibrium can be achieved. Each player acts according to a strategy $u_{n,k} = \phi_{n,k}(x_k)$, and all players' strategies except for player $n$ is denoted $\phi_{-n, k}(\cdot)$.

\begin{definition} \label{def:fne}
{\it
  \textbf{(Local) feedback Nash equilibrium} \\
  A collection of feedback policies $u_{n,k} =
  \phi_{n,k}^\star(x_k)$ is said to be a feedback Nash equilibrium (FNE) to the full game if no player can benefit from changing their policy unilaterally for any subgame, i.e.,
  \begin{align}
      \label{eq:fne}
      J_{n,t}(x_t, \phi^\star_{:,t:}) \leq J_{n,t}(x_t, [\phi_{n,t:}, \phi^\star_{-n, t:}]), \ \forall t \in \{0, 1, ..., T\}
  \end{align}
  where $J_{n,t}(x_t, \phi_{:, t:})$ indicates the total cost of player
  $n$ when all players follow policy $\phi_{:,t:}$ for subgame
  $t$.

  Furthermore, the FNE is local around $\bar u_{n,k} = \phi_{n,k}^*(\bar x_k)$ $\forall n \in \{1, 2, ..., N\}, \forall k \in \{0, 1, \dots, T-1\}$, if \eqref{eq:fne} holds only locally and the resulting trajectories remain in a neighborhood of $[\bar x,\bar u]$.
}
\end{definition}

Ideally, feedback Nash equilibrium can be solved via Bellman recursion, which originated from optimal control and was extended to dynamic games \cite{haurie2012games, krawczyk2018multistage}. Instead of solving for the minimizing action at each stage, equilibria of stage-wise games are computed via the following recursion:
\begin{subequations}
  \label{eq:bellman}
  \begin{align}
    V_{n,T+1}^\star(x_{T+1}) &= 0 \\
    Q_{n,k}^\star(x_k,u_{:,k}) &= c_{n,k}(x_k,u_{:,k}) +
                                 V_{n,k+1}^\star(f_k(x_k,u_{:,k})) \\
    \label{eq:QGame}
    V_{n,k}^\star(x_k) &= \min_{u_{n,k}} Q_{n,k}^\star(x_k,u_{:,k}) \\
    k &= 0, 1, ..., T
  \end{align}
\end{subequations}
Here $V^{\star}_{n, k}(x_k)$ and $Q^{\star}_{n, k}(x_k, u_{:, k})$
are referred to as \textit{equilibrium value functions} for player $n$
at time step $k$. In particular, if a solution to the Bellman recursion is
found, the corresponding optimal strategy for player $n$ at time $k$
would be the $u_{n,k}$ which minimizes
$Q_{n,k}^\star(x_k,u_{:,k})$. Note that \eqref{eq:QGame} defines a
static game with respect to the $u_{:,k}$ variable at step $k$.
A well known verification theorem states that a feedback policy
$u_{:,k} = \phi_{k}^\star(x_k), \ k = 0, 1, ..., T$ solving the
sequence of static games defined by \eqref{eq:QGame}, is a subgame
perfect FNE for the dynamic game \cite{krawczyk2018multistage,
  haurie2012games}. For general dynamic games, the Bellman equations
are not computationally tractable. Note that the game ends at $k = T$, and setting $V_{n,T+1}^\star(x_{T+1}) = 0$ is only for ease of describing the Bellman recursion.

\subsection{Existence of Solutions and Convergence Conditions}
\label{sec:existence}
To guarantee convergence, we assume that $\J(u)$ satisfies the
smoothness and non-degeneracy conditions required by Newton's method \cite{nocedal2006numerical}.
\begin{assumption}[Smoothness]
  \label{asm:smooth}
  The vector-valued function, $\pazocal{J}(u)$, is differentiable with
  locally Lipschitz derivatives.
\end{assumption}
\begin{assumption}[Non-degeneracy]
  \label{asm:inv}
  The Jacobian $\frac{\partial
  \pazocal{J}(u^\star)}{\partial u}$ is invertible.
\end{assumption}

A sufficient condition
for the smoothness assumptions is that the functions $f_k$ and
$c_{n,k}$ are twice continuously differentiable with Lipschitz second
derivatives. For either of our methods, we will solve a sequence of stagewise
quadratic games. As we will see, a sufficient condition for invertibility of
$\frac{\partial \J(u^\star)}{\partial u}$ is the unique solvability of the
stagewise games near the equilibrium.

The non-degeneracy and smoothness conditions guarantee
that Newton's  method converges locally to a stationary point
satisfying \eqref{eq:necessary}. The following assumption guarantees
that this stationary point is a strict local Nash equilibrium.

\begin{assumption}
  \label{asm:convex}
  Each player's Hessian,
  $\frac{\partial^2 J_{n,0}(u^\star)}{\partial^2 u_{n,:}}$, is positive
  definite, i.e. each player's cost $J_n(u)$ is strictly convex w.r.t. their actions $u_{n,:}$.
\end{assumption}

\subsection{Notations of Derivatives}
\label{sec:derivatives}
We define the following shorthand notations for first and second order derivatives of both the dynamics and cost functions for given trajectory $\bar x, \bar u$, which are used in both the stagewise Newton's method and DDP method. The derivatives show up because we are using quadratic approximations around trajectory.
\begin{subequations}
  \label{eq:approximations}
  \begin{align}
    & A_k = \frac{\partial f_k(x_k, u_{:,k})}{\partial x_k} \Big|_{\bar x, \bar u} \quad
      \quad B_{k} = \frac{\partial f_k(x_k, u_{:,k})}{\partial u_{:,k}} \Big|_{\bar x, \bar u} \\
    & G_k^l =
      \begin{bmatrix}
        \frac{\partial^2 f^l_k}{\partial x_k^2} & \frac{\partial^2 f^l_k}{\partial x_k   \partial u_{:,k}} \\
        \frac{\partial^2 f^l_k}{\partial u_{:,k} \partial x_k} & \frac{\partial^2 f^l_k}{\partial u_{:,k}^2}
      \end{bmatrix} \Bigg|_{\bar x, \bar{u}}, \ \  l = 1, 2, \ldots, n_x \\
    & R_k(\delta x_k, \delta u_{:,k}) =
      \begin{bmatrix}
        \begin{bmatrix}
          \delta x_k \\
          \delta u_{:,k}
        \end{bmatrix}^\top G_k^1
        \begin{bmatrix}
          \delta x_k \\
          \delta u_{:,k}
        \end{bmatrix} \\
        \vdots \\
        \begin{bmatrix}
          \delta x_k \\
          \delta u_{:,k}
        \end{bmatrix}^\top G_k^{n_x}
        \begin{bmatrix}
          \delta x_k \\
          \delta u_{:,k}
        \end{bmatrix}
      \end{bmatrix} \\
    & M_{n, k} = \left.
               \begin{bmatrix}
                 2 c_{n, k} & \frac{\partial c_{n, k}}{\partial x_k}
                 & \frac{\partial c_{n, k}}{\partial u_{:,k}} \\
                 \frac{\partial c_{n, k}}{\partial x_k}^\top
                 & \frac{\partial^2 c_{n, k}}{\partial x_k^2}
                 & \frac{\partial^2 c_{n, k}}{\partial x_k \partial u_{:,k}} \\
                 \frac{\partial c_{n, k}}{\partial u_{:,k}}^\top
                 & \frac{\partial^2 c_{n, k}}{\partial u_{:,k}\partial x_k}
                 & \frac{\partial^2 c_{n, k}}{\partial u_{:,k}^2}
               \end{bmatrix} \right\rvert_{\bar x, \bar u}
                   \label{eq:cost_devs} \\
    \nonumber
             & \quad \ \ =
               \begin{bmatrix}
                 M_{n, k}^{11} & M_{n, k}^{1x} & M_{n, k}^{1u} \\
                 M_{n, k}^{x1} & M_{n, k}^{xx} & M_{n, k}^{xu} \\
                 M_{n, k}^{u1} & M_{n, k}^{ux} & M_{n, k}^{uu}
               \end{bmatrix}.
  \end{align}
\end{subequations}

\section{Stagewise Newton's Method}
\label{sec:alg_nm}
This section describes the stagewise Newton's method for dynamic games of the form in Problem~\ref{problem:original}.
Subsection~\ref{sec:algOverview_nm} gives a high-level description of the
algorithms, while Subsection~\ref{sec:algMatrices_nm} describes the explicit matrix calculations.

\subsection{Algorithm Overview}\label{sec:algOverview_nm}
With a given trajectory $\bar u$, the Newton step for solving \eqref{eq:necessary} is given by:
\begin{align}
  \label{eq:direct_newton}
  \frac{\partial \pazocal{J}(\bar u)}{\partial u} \delta u^N = - \pazocal{J}(\bar u).
\end{align}

This rule leads to a quadratic convergence to a root in
\eqref{eq:necessary} whenever $\nabla_u
\pazocal{J}(u)$ is locally Lipschitz and invertible
\cite{nocedal2006numerical}. The next two lemmas give game-theoretic
interpretations of the Newton step.

\begin{lemma} \label{lem:newtonIsQuadApprox}
  {\it
    If Assumptions~\ref{asm:smooth}-~\ref{asm:convex} hold, then solving \eqref{eq:direct_newton} is equivalent to solving the
    quadratic game defined by:
    \begin{align}
      \label{eq:direct_game}
      \min_{\delta u_{n,:}} \quad J_n(\bar u) +
      \frac{\partial J_n(\bar u)}{\partial u} \delta u
      + \frac{1}{2} \delta u^\top \frac{\partial^2 J_n(\bar u)}{\partial u^2} \delta u
    \end{align}
  }
\end{lemma}

\begin{pf*}{Proof.}
  Under the strict local equilibrium assumptions,
  \eqref{eq:direct_game} has a unique solution which is found by
  differentiating with respect to $\delta u_{n,:}$ and setting the
  result to $0$. Stacking these equations leads precisely to
  \eqref{eq:direct_newton}.
  \hfill\qed
\end{pf*}

Throughout the paper, we will assume that
Assumptions~\ref{asm:smooth}-\ref{asm:convex} hold.

The next lemma shows that \eqref{eq:direct_game} can be expressed as a
quadratic dynamic game. It is proved in Appendix~\ref{app:newton_game}.

\begin{lemma}
  \label{lem:newton_game}
  {\it
    The quadratic game defined in \eqref{eq:direct_game} is equivalent
    to the dynamic game defined by:
    \begin{subequations}
      \label{eq:newton_dp}
      \begin{align}
        \label{eq:newton_dp_objective}
        & \min_{u_{n,:}}\frac{1}{2} \sum_{k=0}^T \left(
          \begin{bmatrix}
            1 \\
            \delta x_k \\
            \delta u_{:,k}
          \end{bmatrix}^\top
        M_{n, k}
        \begin{bmatrix}
          1 \\
          \delta x_k \\
          \delta u_{:,k}
        \end{bmatrix}
        + M^{1x}_{n,k} \Delta x_k \right)
      \end{align}
      subject to
      \begin{align}
        & \quad \quad \delta x_0 = 0 \\
        \label{eq:newton_dp_init1}
        & \quad \quad \Delta x_0 = 0 \\
        \label{eq:newton_dp_dynamics0}
        & \quad \quad \delta x_{k+1} = A_k \delta{x}_k + B_k \delta u_{:,k} \\
        \label{eq:newton_dp_dynamics1}
        & \quad \quad \Delta x_{k+1} = A_k \Delta x_k + R_k(\delta x_k, \delta u_{:,k})\\
        & \quad \quad k = 0, 1, \ldots, T
      \end{align}
    \end{subequations}
  }
  where $A_k$, $B_k$, $M_{n,k}$, $R_k(\delta x_k, \delta u_{:,k})$ are defined in Section~\ref{sec:derivatives}, which are constants for given trajectory $\bar u$.
\end{lemma}

Note that the states of the dynamic game are given by $\delta x_k$ and
$\Delta x_k$ as
\begin{subequations}
  \label{eq:newton_dp_states}
  \begin{align}
      \label{eq:newton_dp_states0}
      & \delta x_k = \sum_{i = 0}^{T} \frac{\partial x_k}{\partial u_{:,i}} \Big|_{\bar x, \bar{u}} \delta u_{:,i} \\
      \label{eq:newton_dp_states1}
      & \Delta x_k^l = \sum_{i = 0}^{T} \sum_{j = 0}^{T} \delta u_{:,i}^\top {\frac{\partial^2 x_k^l}{\partial u_{:,i} \partial u_{:,j}}} \Big|_{\bar x,\bar{u}} \delta u_{:,j}, \  l = 1, 2, \ldots, n_x
  \end{align}
\end{subequations}

It turns out that the Bellman equations \eqref{eq:bellman} associated
with problem \eqref{eq:newton_dp} can be solved analytically and the
resulting value functions have quadratic forms. The next lemma
describes the explicit solution to \eqref{eq:newton_dp} based on
Bellman equation \eqref{eq:bellman}. It is proved in Appendix
\ref{app:newtonMatrices}.

\begin{lemma}
  \label{lem:newtonMatrices}
  {\it
    The equilibrium value functions for the dynamic game defined by \eqref{eq:newton_dp} are denoted as $\hat {V}^{\bar u}_{n,k}(\cdot)$ and $\hat Q^{\bar u}_{n,k}(\cdot, \cdot)$, which can be expressed as
    \begin{subequations}
    \label{eq:newton_dp_solution_val_fun}
      \begin{align}
        \label{eq:newton_dp_solution_state_val_fun}
        & \hat V^{\bar u}_{n,k}(\delta x_k, \Delta x_k) = \frac{1}{2} \left(
          \begin{bmatrix}
            1 \\
            \delta x_k
          \end{bmatrix}^\top S_{n,k}
        \begin{bmatrix}
          1 \\
          \delta x_k
        \end{bmatrix} + \Omega_{n,k} \Delta x_k \right)
        \\
        \label{eq:newton_dp_solution_state_action_val_fun}
        & \hat Q^{\bar u}_{n,k}(\delta x_k, \Delta x_k, \delta u_{:,k}) = \nonumber \\
        & \quad \quad \quad \quad \quad \ \frac{1}{2} \left(
          \begin{bmatrix}
            1 \\
            \delta x_k \\
            \delta u_{:,k}
          \end{bmatrix}^\top \Gamma_{n,k}
        \begin{bmatrix}
          1 \\
          \delta x_k \\
          \delta u_{:,k}
        \end{bmatrix} + \Omega_{n,k} \Delta x_k \right)
      \end{align}
    \end{subequations}
    where the matrices $S_{n,k}$, $\Gamma_{n,k}$, and $\Omega_{n,k}$
    can be computed in a backward pass. Note that we use the superscript $^{\bar u}$ to indicate the nominal trajectory that we are approximating the original problem around. Detailed descriptions are given by \eqref{eq:newton_dp_matrices} in Section~\ref{sec:algMatrices_nm}.
  }
\end{lemma}

The next lemma gives the form of the solution based on the value functions. Note that \eqref{eq:newton_dp_solution_state_action_val_fun} is now a quadratic game in the $u_{:,k}$ variables which has unique solution \cite{basar1999dynamic}.
The solvability of these stagewise games indicates that the dynamic game \eqref{eq:newton_dp} is solvable, hence the equivalent game \eqref{eq:direct_newton} has a solution and $\frac{\partial \J(u^\star)}{\partial u}$ is invertible as we alluded to in Section~\ref{sec:existence}.

A sufficient condition for solvability of these games is
given in terms of $\J(u)$ is given in the following lemma.
Its proof is in Appendix~\ref{app:nmSol}.

\begin{lemma}
  \label{lem:nmSol}
  {\it
    If $\frac{\partial \J(\bar u)}{\partial u}$ is invertible, the game defined by
    \eqref{eq:newton_dp} has a unique solution of the form:
    \begin{equation}
      \label{eq:nm_localSolution}
      u_{:,k} = \bar u_{:,k} +  K_k \delta x_k + s_k.
    \end{equation}
  }
\end{lemma}

\subsection{Details of Stagewise Newton's Method}
\label{sec:algMatrices_nm}
\begin{lemma}
\label{lem:nmMatrices}
\textit
  The matrices $S_{n,k}$, $\Gamma_{n,k}$, and $\Omega_{n,k}$ in \eqref{eq:newton_dp_solution_val_fun} are computed recursively by $S_{n,T+1}=0$, $\Omega_{n,T+1} = 0$, and
  \begin{subequations}
    \label{eq:newton_dp_matrices}
    \begin{align}
      \label{eq:newton_dp_solution_omega1}
      & \Omega_{n,k} = M_{n,k}^{1x} + \Omega_{n,k+1} A_k \\
      \label{eq:newton_dp_solution_D}
      & D_{n,k} = \sum_{l=1}^{n_x} \Omega_{n,k+1}^l G_k^l  \\
      \label{eq:GammaDef}
      & \Gamma_{n,k} = M_{n,k} \nonumber \\
      & \quad \quad \ \ +
        \begin{bmatrix}
          S_{n, k+1}^{11} & S_{n,k+1}^{1x}A_k & S_{n,k+1}^{1x}B_k \\
          A_k^\top S_{n,k+1}^{x1} & A_k^\top S_{n,k+1}^{xx}A_k + D_k^{xx} & A_k^\top S_{n,k+1}^{xx} B_k + D_k^{xu} \\
          B_k^\top  S_{n,k+1}^{x1}  & B_k^\top  S_{n,k+1}^{xx}A_k + D_k^{ux} & B_k^\top S_{n,k+1}^{xx} B_k + D_k^{uu}
        \end{bmatrix} \\
      & \quad \ =
        \begin{bmatrix}
          \Gamma_{n, k}^{11} & \Gamma_{n, k}^{1x} & \Gamma_{n, k}^{1u_1} & \Gamma_{n, k}^{1u_2} & \cdots & \Gamma_{n, k}^{1u_N} \\
          \Gamma_{n, k}^{x1} & \Gamma_{n, k}^{xx} & \Gamma_{n, k}^{xu_1} & \Gamma_{n, k}^{xu_2} & \cdots & \Gamma_{n, k}^{xu_N} \\
          \Gamma_{n, k}^{u_11} & \Gamma_{n, k}^{u_1x} & \Gamma_{n, k}^{u_1u_1} & \Gamma_{n, k}^{u_1u_2} & \cdots & \Gamma_{n, k}^{u_1u_N} \\
          \Gamma_{n, k}^{u_21} & \Gamma_{n, k}^{u_2x} & \Gamma_{n, k}^{u_2u_1} & \Gamma_{n, k}^{u_2u_2} & \cdots & \Gamma_{n, k}^{u_2u_N} \\
          \vdots & \vdots & \vdots & \vdots & \ddots & \vdots \\
          \Gamma_{n, k}^{u_N1} & \Gamma_{n, k}^{u_Nx} & \Gamma_{n, k}^{u_Nu_1} & \Gamma_{n, k}^{u_Nu_2} & \cdots & \Gamma_{n, k}^{u_Nu_N}
        \end{bmatrix} \\
      \label{eq:newton_dp_invert}
      & F_k =
        \begin{bmatrix}
          \Gamma_{1k}^{u_1u} \\
          \Gamma_{2k}^{u_2u} \\
          \vdots \\
          \Gamma_{Nk}^{u_Nu}
        \end{bmatrix} =
      \begin{bmatrix}
        \Gamma_{1k}^{u_1u_1} & \Gamma_{1k}^{u_1u_2} & \cdots & \Gamma_{1k}^{u_1u_N} \\
        \Gamma_{2k}^{u_2u_1} & \Gamma_{2k}^{u_2u_2} & \cdots & \Gamma_{2k}^{u_2u_N} \\
        \vdots & \vdots & \ddots & \vdots \\
        \Gamma_{Nk}^{u_Nu_1} & \Gamma_{Nk}^{u_Nu_2} & \cdots & \Gamma_{Nk}^{u_Nu_N}
      \end{bmatrix} \\
      & P_k =
        \begin{bmatrix}
          \Gamma_{1k}^{u_1x} \\
          \Gamma_{2k}^{u_2x} \\
          \vdots \\
          \Gamma_{Nk}^{u_Nx}
        \end{bmatrix}, \quad
      H_k =
      \begin{bmatrix}
        \Gamma_{1k}^{u_11} \\
        \Gamma_{2k}^{u_21} \\
        \vdots \\
        \Gamma_{Nk}^{u_N1}
      \end{bmatrix} \\
      \label{eq:newton_strategy}
      & s_k = - F_k^{-1} H_k, \quad K_k = - F_k^{-1} P_k \\
      \label{eq:newton_dp_solution_S}
      & S_{n, k} =
        \begin{bmatrix}
          1 & 0 & s_k^\top \\
          0 & I & K_k^\top
        \end{bmatrix} \Gamma_{n, k}
                         \begin{bmatrix}
                           1 & 0 \\
                           0 & I \\
                           s_k & K_k
                         \end{bmatrix}
    \end{align}
  \end{subequations}
  for $k = T,T-1,\ldots,0$.
\end{lemma}

\begin{pf*}{Proof.}
  By construction we must have $S_{n,T+1}=0$. Plugging \eqref{eq:newton_dp_dynamics0} and \eqref{eq:newton_dp_dynamics1} into \eqref{eq:newton_dp_solution_state_val_fun} gives the backward iteration of \eqref{eq:newton_dp_solution_omega1}\eqref{eq:newton_dp_solution_D}\eqref{eq:GammaDef}. Since $u_{:,k} = \bar u_{:,k}+\delta u_{:,k}$ and $\bar u_{:,k}$ is
  constant, the static game defined in \eqref{eq:newton_dp_solution_state_action_val_fun} can be
  solved in the $\delta u_{:,k}$ variables. Differentiating
  \eqref{eq:newton_dp_solution_state_action_val_fun} by $\delta u_{n,k}$, collecting the derivatives for all players and setting them to zero leads to the
  necessary condition for an equilibrium:
  \begin{equation}
    \label{eq:nm_necessary}
    F_k \delta u_{:,k} + P_k \delta x_k + H_k = 0.
  \end{equation}
  Thus, the matrices for the equilibrium strategy are given in
  \eqref{eq:newton_strategy}. Plugging \eqref{eq:nm_localSolution} into
  \eqref{eq:newton_dp_solution_state_action_val_fun} leads to \eqref{eq:newton_dp_solution_S}.
  \hfill\qed
\end{pf*}

\section{DDP Algorithms for Dynamic Games}
\label{sec:alg_ddp}

This section describes the differential dynamic programming algorithm for dynamic games of the form in Problem~\ref{problem:original}.
Subsection~\ref{sec:algOverview_ddp} gives a high-level description of the
algorithms, while Subsection~\ref{sec:algMatrices_ddp} describe the explicit matrix calculations.

\subsection{Algorithm Overview}
\label{sec:algOverview_ddp}
The idea of the differentiable dynamic programming (DDP) is to
maintain quadratic approximations of $V_{n,k}^*$ and $Q_{n,k}^*$
around a trajectory $\bar u$ denoted by $\tilde V^{\bar u}_{n,k}$ and $\tilde Q^{\bar u}_{n,k}$, respectively.

We need some notation for our approximations.
For a scalar-valued function, $h(z)$, we denote the quadratic
approximation near $\bar{z}$ by:
\begin{subequations}
  \label{eq:qApprox}
  \begin{align}
    \qApprox(h(z))_{\bar z} =&\frac{1}{2}
                               \begin{bmatrix}
                                 1 \\
                                 \delta z
                               \end{bmatrix}^\top
    \begin{bmatrix}
      2 h(\bar z) & \frac{\partial h(\bar z)}{\partial z} \\
      \frac{\partial h(\bar z)}{\partial z}^\top & \frac{\partial^2
        h(\bar z)}{\partial z^2}
    \end{bmatrix}
                                                   \begin{bmatrix}
                                                     1 \\
                                                     \delta z
                                                   \end{bmatrix}
    \\
    \delta z =& z-\bar z.
  \end{align}
\end{subequations}
If $h : \R^n\to \R^m$ we form the quadratic approximation by stacking
all of the quadratic approximations of the entries:
\begin{equation}
  \label{eq:stackedQuad}
  \qApprox(h(z))_{\bar z} = [\qApprox(h_1(z))_{\bar
    z}, \ \ldots,\ \qApprox(h_m(z))_{\bar z}]^\top
\end{equation}

Let $\bar x_k$ and $\bar
u_{:,k}$ be a trajectory of states and actions satisfying the dynamic
equations from \eqref{eq:dynamics} and $z_k = [x_k^\top, u_{:,k}^\top]^\top$.
The approximate Bellman recursion around this trajectory is given by:
\begin{subequations}
  \label{eq:bellmanApprox}
  \begin{align}
    \tilde V^{\bar u}_{n,T+1}(x_{T+1}) &= 0 \\
    \label{eq:Qquad}
    \tilde Q^{\bar u}_{n,k}(z_k) &= \qApprox(
                          c_{n,k}(z_k) + \tilde V^{\bar u}_{n,k+1}(f_k(z_k))
                          )_{\bar z_k}\\
    \label{eq:quadQGame}
    \tilde V^{\bar u}_{n,k}(x_k) &= \min_{u_{n,k}} \tilde Q^{\bar u}_{n,k}(x_k,u_{:,k}).
  \end{align}
\end{subequations}

The quadratic approximation is possible because $f_{k}(z_k)$ and $c_{n,k}(z_k)$ are twice differentiable. Similar to stagewise Newton's method and Lemma~\ref{lem:nmSol}, the following lemma describes the form of solution to \eqref{eq:quadQGame}. It is proven in Appendix~\ref{app:ddpSol}.

\begin{lemma}
  \label{lem:quadSol}
  {\it
    If $\frac{\partial \J(\bar u)}{\partial u}$ is invertible, the game defined by
    \eqref{eq:quadQGame} has a unique solution of the form:
    \begin{equation}
      \label{eq:localSolution}
      u_{:,k} = \bar u_{:,k} +  \tilde K_k \delta x_k + \tilde s_k.
    \end{equation}
  }
\end{lemma}

In the notation defined above, we have
that $\delta x_k = x_k - \bar x_k$.
Note that if
$\frac{\partial \J(u^\star)}{\partial u}$ is invertible, then
$\frac{\partial \J(\bar
  u)}{\partial u}$ is invertible for all $\bar u$ in a neighborhood of $u^\star$.

Here we provide the pseudo code for both algorithms. Note that an initial trajectory of $\bar x$ should be found by running the system with actions $\bar u$, which are needed to compute derivatives in \eqref{eq:approximations}.
\begin{algorithm}
  \caption{\label{alg:nm_ddp} Stagewise Newton's and DDP methods for Nonlinear Dynamic Games}
  \begin{algorithmic}
    \State Generate an initial trajectory $\bar x, \bar u$
    \Loop
    \State{\textbf{Backward Pass:}}
      \If{Newton's method}
        \State Form the approximated dynamic game \eqref{eq:newton_dp}
        \State Compute $K_k$ and $s_k$ from \eqref{eq:nm_localSolution}.
      \EndIf
      \If{DDP}
        \State Perform approximated Bellman recursion \eqref{eq:bellmanApprox}
        \State Compute $\tilde K_k$ and $\tilde s_k$ from \eqref{eq:localSolution}.
      \EndIf
    \State{\textbf{Forward Pass:}}
    \State{Generate a new trajectory using the affine policy defined
      by $K_k, s_k$ or $\tilde K_k, \tilde s_k$}
      \State Check convergence
    \EndLoop
    \State Obtaining stationary trajectory $u^{\star}$, $x^{\star}$ and feedback policy $\tilde K_k^\star$, $\tilde s_k^\star$
  \end{algorithmic}
\end{algorithm}

\subsection{Details of DDP method}
\label{sec:algMatrices_ddp}

Using the notation from \eqref{eq:approximations}, \eqref{eq:qApprox}, \eqref{eq:stackedQuad} and $z_k$, the second-order approximations of the dynamics and cost are given by:
\begin{subequations}
  \begin{align}
    \label{eq:dynQuad}
    & \qApprox(f_k(z_k))_{\bar z_k} = f_k(\bar z_k) +  A_k \delta x_k +
                                    B_k \delta u_{:,k} + R_k(\delta z_k) \\
    & \qApprox(c_{n,k}(z_k))_{\bar z_k} = \begin{bmatrix}
      1 \\
      \delta z_k
    \end{bmatrix}^\top M_{n,k}
    \begin{bmatrix}
      1 \\
      \delta z_k
    \end{bmatrix}.
  \end{align}
\end{subequations}

By construction $\tilde V^{\bar u}_{n,k}(x_k)$ and $\tilde Q^{\bar u}_{n,k}(x_k,u_{:,k})$ are
quadratic, so there must be matrices $\tilde S_{n,k}$ and $\tilde
\Gamma_{n,k}$ such that
\begin{subequations}
  \label{eq:ddp_cost_matrices}
  \begin{align}
    \label{eq:ddpV}
    \tilde V^{\bar u}_{n,k}(x_k) &= \frac{1}{2}
                          \begin{bmatrix}
                            1 \\
                            \delta x_k \\
                          \end{bmatrix}^\top \tilde S_{n, k}
                                                   \begin{bmatrix}
                                                     1 \\
                                                     \delta x_k \\
                                                   \end{bmatrix} \\
    \label{eq:ddpQ}
    \tilde Q^{\bar u}_{n,k}(x_k, u_{:,k}) &= \frac{1}{2}
                                   \begin{bmatrix}
                                     1 \\
                                     \delta x_k \\
                                     \delta u_{:,k}
                                   \end{bmatrix}^\top \tilde\Gamma_{n, k}
    \begin{bmatrix}
      1 \\
      \delta x_k \\
      \delta u_{:,k}
    \end{bmatrix}.
  \end{align}
\end{subequations}

\begin{lemma}
  \label{lem:ddp_matrices}
  {\it
    The matrices in \eqref{eq:ddp_cost_matrices} are defined recursively
    by $\tilde S_{n,T+1} = 0$ and:
    \begin{subequations}
      \label{eq:ddp_backward_pass}
      \begin{align}
        \label{eq:TildeDDef}
        & \tilde D_{n,k} = \sum_{l=1}^{n_x} \tilde S_{n,k+1}^{1x^l}
          G_k^l
        \\
        & \tilde \Gamma_{n, k} = M_{n, k} \nonumber \\
        \label{eq:TildeGammaBackprop}
        &+
          \begin{bmatrix}
            \tilde S_{n,k+1}^{11} & \tilde S_{n,k+1}^{1x}A_k & \tilde S_{n,k+1}^{1x}B_k \\
            A_k^\top \tilde S_{n,k+1}^{x1} & A_k^\top \tilde S_{n,k+1}^{xx}A_k + \tilde D_{n, k}^{xx} & A_k^\top \tilde S_{n,k+1}^{xx} B_k + \tilde D_{n, k}^{xu} \\
            B_k^\top \tilde S_{n,k+1}^{x1}  & B_k^\top \tilde S_{n,k+1}^{xx}A_k + \tilde D_{n,k}^{ux} & B_k^\top \tilde S_{n,k+1}^{xx} B_k + \tilde D_{n, k}^{uu}
          \end{bmatrix} \\
        &=
          \begin{bmatrix}
            \tilde \Gamma_{n, k}^{11} & \tilde \Gamma_{n, k}^{1x} & \tilde \Gamma_{n, k}^{1u_1} & \tilde \Gamma_{n, k}^{1u_2} & \cdots & \tilde \Gamma_{n, k}^{1u_N} \\
            \tilde \Gamma_{n, k}^{x1} & \tilde \Gamma_{n, k}^{xx} & \tilde \Gamma_{n, k}^{xu_1} & \tilde \Gamma_{n, k}^{xu_2} & \cdots & \tilde \Gamma_{n, k}^{xu_N} \\
            \tilde \Gamma_{n, k}^{u_11} & \tilde \Gamma_{n, k}^{u_1x} & \tilde \Gamma_{n, k}^{u_1u_1} & \tilde \Gamma_{n, k}^{u_1u_2} & \cdots & \tilde \Gamma_{n, k}^{u_1u_N} \\
            \tilde \Gamma_{n, k}^{u_21} & \tilde \Gamma_{n, k}^{u_2x} & \tilde \Gamma_{n, k}^{u_2u_1} & \tilde \Gamma_{n, k}^{u_2u_2} & \cdots & \tilde \Gamma_{n, k}^{u_2u_N} \\
            \vdots & \vdots & \vdots & \vdots & \ddots & \vdots \\
            \tilde \Gamma_{n, k}^{u_N1} & \tilde \Gamma_{n, k}^{u_Nx} & \tilde \Gamma_{n, k}^{u_Nu_1} & \tilde \Gamma_{n, k}^{u_Nu_2} & \cdots & \tilde \Gamma_{n, k}^{u_Nu_N}
          \end{bmatrix} \\
        \label{eq:TildeFDef}
        & \tilde F_k =
          \begin{bmatrix}
            \tilde \Gamma_{1,k}^{u_1u} \\
            \tilde \Gamma_{2,k}^{u_2u} \\
            \vdots \\
            \tilde \Gamma_{N,k}^{u_Nu}
          \end{bmatrix} =
        \begin{bmatrix}
          \tilde \Gamma_{1,k}^{u_1u_1} & \tilde \Gamma_{1,k}^{u_1u_2} & \cdots & \tilde \Gamma_{1,k}^{u_1u_N} \\
          \tilde \Gamma_{2,k}^{u_2u_1} & \tilde \Gamma_{2,k}^{u_2u_2} & \cdots & \tilde \Gamma_{2,k}^{u_2u_N} \\
          \vdots & \vdots & \ddots & \vdots \\
          \tilde \Gamma_{N,k}^{u_Nu_1} & \tilde \Gamma_{N,k}^{u_Nu_2} & \cdots & \tilde \Gamma_{N,k}^{u_Nu_N}
        \end{bmatrix} \\
        \label{eq:TildePHsKDef}
        & \tilde P_k =
          \begin{bmatrix}
            \tilde \Gamma_{1,k}^{u_1x} \\
            \tilde \Gamma_{2,k}^{u_2x} \\
            \vdots \\
            \tilde \Gamma_{N,k}^{u_Nx}
          \end{bmatrix}, \quad
        \tilde H_k =
        \begin{bmatrix}
          \tilde \Gamma_{1,k}^{u_11} \\
          \tilde \Gamma_{2,k}^{u_21} \\
          \vdots \\
          \tilde \Gamma_{N,k}^{u_N1}
        \end{bmatrix} \\
        \label{eq:ddpStrategyMatrices}
        & \tilde s_k = - \tilde F_k^{-1} \tilde H_k, \quad \tilde K_k =
          - \tilde F_k^{-1} \tilde P_k \\
        \label{eq:TildeSDef}
        & \tilde S_{n, k} =
          \begin{bmatrix}
            1 & 0 & \tilde s_k^\top \\
            0 & I  & \tilde K_k^\top
          \end{bmatrix} \tilde \Gamma_{n, k}
                                   \begin{bmatrix}
                                     1 & 0 \\
                                     0 & I \\
                                     \tilde s_k & \tilde K_k
                                   \end{bmatrix} ,
      \end{align}
    \end{subequations}
    for $k=T,T-1,\ldots,0$.
  }
\end{lemma}

\begin{pf*}{Proof.}
  By construction we must have $\tilde S_{n,T+1}=0$. Plugging \eqref{eq:dynQuad} into \eqref{eq:Qquad}
  and dropping all cubic and higher terms gives \eqref{eq:TildeDDef}\eqref{eq:TildeGammaBackprop}. Since
  $u_{:,k} = \bar u_{:,k}+\delta u_{:,k}$ and $\bar u_{:,k}$ is
  constant, the static game defined in \eqref{eq:quadQGame} can be
  solved in the $\delta u_{:,k}$ variables. Differentiating
  \eqref{eq:ddpQ} by $\delta u_{n,k}$, collecting the derivatives for all players and setting them to zero leads to the
  necessary condition for an equilibrium:
  \begin{equation}
    \label{eq:ddp_necessary}
    \tilde F_k \delta u_{:,k} + \tilde P_k \delta x_k + \tilde H_k = 0.
  \end{equation}
  Thus, the matrices for the equilibrium strategy are given in
  \eqref{eq:ddpStrategyMatrices}. Plugging \eqref{eq:localSolution} into
  \eqref{eq:ddpQ} leads to \eqref{eq:TildeSDef}.
  \hfill\qed
\end{pf*}

We can see that the matrices used in the recursions
for both DDP and stagewise Newton's method are very similar in
structure. Indeed, the iterations are identical aside from the
definitions of the $D_{n,k}$ and $\tilde D_{n,k}$ matrices.

\section{Convergence and Equilibria}
\label{sec:convergence}

\begin{remark}
  The calculations of Newton's method and DDP are for games are
  similar to those arising in single-agent optimal control.
  The difference is that the game case inverts the matrices $F_k$ and
  $\tilde F_k$  which are constructed from submatrices of the
  value function matrices, $\Gamma_{n,k}$ and $\tilde \Gamma_{n,k}$. In
  contrast, the single agent algorithms invert $\Gamma_{n,k}$ and $\tilde
  \Gamma_{n,k}$ directly.
  It is due to this difference, that the proof for game scenario
requires separate though similar treatment to those of
\cite{murray1984differential, dunn1989efficient}.
\end{remark}

Throughout this section we will assume that both methods are starting from the same initial action trajectory $\bar
u$ that is close to the stationary point $u^{\star}$ such that $\norm{\bar u - u^{\star}} = \epsilon$. Let $u^N$ and $u^D$ be the updated action trajectories of stagewise Newton's
method and DDP, respectively. We define update steps $\delta u^N$ and
$\delta u^D$
\begin{equation}
  u^N = \bar u + \delta u^N \quad u^D = \bar u + \delta u^D.
\end{equation}

Now we are ready to introduce Theorem~\ref{thm:main}, which is our first main result.

\begin{theorem}
  \label{thm:main}
  {\it
    If $u^\star$ satisfies the necessary conditions \eqref{eq:necessary} and that
    $\frac{\partial \J(u^*)}{\partial u}$ is invertible, then both the stagewise Newton
    and DDP algorithms converge locally to $u^{\star}$ at a quadratic
    rate. Furthermore, if Assumption \ref{asm:convex} is true, $u^\star$ is a time-consistent, open-loop Nash equilibrium.
  }
\end{theorem}

\begin{pf*}{Proof.}
  \label{sec:proof}
  The convergence rate for stagewise Newton is natural since it is exactly Newton's step to the root finding problem of \eqref{eq:direct_newton} \cite{nocedal2006numerical}. Furthermore, the Newton step satisfies:
  \begin{equation}
    \|\bar u + \delta u^N - u^\star\| = O(\epsilon^2).
  \end{equation}
  See~\cite{nocedal2006numerical}.
  DDP method generates an update that is quadratically close to that of stagewise Newton's method, i.e. $\|\delta u^N - \delta u^D\| =
  O(\epsilon^2)$, which is supported by Lemma \ref{lem:matrixClose} and Lemma \ref{lem:solClose} in Appendix \ref{app:closeness_lems}, it inherits the same quadratic convergence rate to local stationary point. The proof is completed by the following steps:
  \begin{subequations}
    \begin{align}
      \|\bar u + \delta u^D - u^{\star}\| &= \| \bar u + \delta u^N - u^{\star} + \delta
                                            u^D - \delta u^N \| \\
                                          &\le \|\bar u + \delta u^N- u^{\star}\| +
                                            \|\delta u^D - \delta u^N\| \\
                                          &= O(\epsilon^2).
    \end{align}
  \end{subequations}
  Note that if Assumption \ref{asm:convex} holds, i.e. each $J_n$ is
  convex with respect to $u_{n,:}$, \eqref{eq:necessary}
  implies that the cost of player $n$ is minimized when the actions of
  the others are fixed, therefore $u^{\star}$ is an open-loop Nash equilibrium. By definition, an OLNE $u^{\star}$ for unconstrained games is time-consistent for dynamic games \cite{krawczyk2018multistage}
  \hfill\qed
\end{pf*}

Next we study the two closed-loop policies found by the stagewise Newton $u_{:,k} = \hat \phi^{\star}_k(x_k) = u_{:,k}^{\star} + K_k^{\star} \delta x_k + s_k^{\star}$ and DDP $u_{:,k} = \tilde \phi^{\star}_k(x_k) =u_{:,k}^{\star} + \tilde K_k^{\star} \delta x_k + \tilde s_k^{\star}$.
Our second theorem states that the feedback policies generated by
stagewise Newton and DDP are approximate local feedback Nash equilibria. The proof is given in Appendix \ref{app:fne}.

\begin{theorem}
  \label{thm:fne}
  The feedback policies by stagewise Newton $\hat \phi^{\star}_k(\cdot)$ and DDP $\tilde \phi^{\star}_k(\cdot)$ are local feedback $O(\epsilon^2)$-Nash equilibria in the sense of Definition~\ref{def:fne}. More specifically,
  \begin{subequations}
    \begin{align*}
      J_{n, t}(x_t, \hat \phi^{\star}_{:, t:}) \leq J_{n, t}(x_t, \phi_{n,t:}, \hat \phi_{-n, t:}^{\star}) + O(\epsilon^2), \  \forall t, n \\
      J_{n, t:}(x_t, \tilde \phi^{\star}_{:, t:}) \leq J_{n, t}(x_t, \phi_{n,t:}, \tilde \phi_{-n, t:}^{\star}) + O(\epsilon^2), \  \forall t, n
    \end{align*}
  \end{subequations}
\end{theorem}

Note that OLNEs are not subgame perfect. Despite this weakness, OLNEs are still valuable in cases where no feedback information is available, a model predictive control (MPC) style strategy is applied or simply the system is sufficiently deterministic. FNEs, on the other hand, are practical for stochastic applications as is as long as the system does not deviate too far from the nominal trajectory.

\section{Implementation Details of the Algorithms}
\label{sec:implementation_details}
Despite their different origins, the two methods are almost the same
for applications. In general, it is hard to tell which method works
better for a specific application beforehand. In cases when there are
multiple equilibrium strategies, the algorithms might converge to
different ones when started from different initial trajectories.

\subsection{Computing Derivatives}
For complicated nonlinear dynamics and costs, modern algorithmic differentiation (AD) software packages, such as Tensorflow \cite{tensorflow2015-whitepaper}, Pytorch \cite{paszke2017automatic},
CasADi \cite{Andersson2018}, are strongly favored.
Section \ref{sec:algMatrices_ddp} provides a form of DDP method based in which derivatives of $c_{n,k}(x_k, u_{:,k})$ and $f_k(x_k, u_{:,k})$. However, with the help of AD software, it is not the only way in practice.
In the backward pass for the DDP method, the quadratic
approximation in \eqref{eq:Qquad} and $\tilde \Gamma_k$ can be directly
computed via automatic differentiation, without computing the derivatives
of a single step $A_k$, $G_k^l$, $M_{n,k}$ in
\eqref{eq:approximations} or keeping $\tilde D_{n,k}$ in
\eqref{eq:TildeDDef}.

\subsection{Regularization}
To ensure that the algorithm converges
regardless of initial condition, a Levenberg\--Marquardt style
regularization should be employed. Such regularization has been used in DDP algorithms for optimal control to
ensure that the required inverses exist and that the solution
improves \cite{liao1991convergence,tassa2011theory}. We found in practice that regularization is essential to the stability for both algorithms. We only use the notations for DDP for simplicity in this section but the same insights hold for stagewise Newton's method. At each step $k$ of the backward pass, we checked the minimal eigenvalue $e_{n,k}$ of matrix $\tilde \Gamma_{n, k}$. If the minimal eigenvalue $e_{n,k}$ is less than a positive value $\lambda$, we reset $\tilde \Gamma_{n, k}$ as
\begin{align}
  \label{eq:regularizeGamma}
  \tilde \Gamma_{n,k} \leftarrow \tilde \Gamma_{n,k} + (\lambda - e_{n,k}) I, \quad n = 1, 2, \cdots, N
\end{align}
where $I$ is an identity matrix. The regularization penalizes
large steps in $\delta x$ and $\delta u$. Thus, it improves the
stability of the algorithm, but sacrifices speed of convergence. The $\lambda$ in our examples are chosen via experimental trials and kept constant for our examples in section \ref{sec:example}. When the algorithm was insufficiently regulated, the trajectories over iterations did not converge from the initial trajectory for our examples. Changing the regularization over iterations while guaranteeing quadratic convergence has been studied in differential dynamic programming literature and is referred to as \textit{adaptive shift for DDP} \cite{liao1991convergence}. Since it is not the focus of this paper, we settled at a constant regularization that enabled smooth and steady improvement of trajectories over iterations for our examples.

\subsection{Computational Complexity}
The computational cost of both algorithms come mainly from evaluating multiple derivatives according to \eqref{eq:approximations} and do backward passes according to either \eqref{eq:newton_dp_matrices} or \eqref{eq:ddp_backward_pass}.
The former requires $T$ evaluations of $A_{k}$, $B_{k}$, which are the first order derivatives of the dynamic, $T n_x$ evaluations of $G_k^l$, which is the second order derivative of the dynamic.
$TN$ evaluations of $M_{n,k}$ are also required as they contain the first and second order derivatives of the cost functions. The latter consists of mainly matrix multiplication and solving linear equations in either \eqref{eq:newton_strategy} or \eqref{eq:ddpStrategyMatrices}.
The overall complexity for both, depending on the implementation, can roughly vary from $O(\min(n_u^2, n_x^2))$ to $O(\max(n_u^3, n_x^3))$ for each step in the backward pass, which is constant for given system and agents.
The advantage of both algorithms over other general GNEP methods is that, the complexity w.r.t. number of stages is linear, i.e. $O(T)$, since the only dependency on it is that we need to do each stagewise computation for $T$ times.
As can be seen based on the complexity analysis, the algorithms are better suited for longer horizon (or finer discretization of continuous problems) dynamic games, rather than games with a large or infinite number of agents.

\section{Numerical Examples}

\label{sec:example}
We apply the proposed algorithms for deterministic nonlinear dynamic games to two examples in this section. We compare the performances of both algorithms on a simple toy example first, and then apply the DDP method to a more complicated problem. We gain proof of concept that both methods performs reasonably close in practice and that they can be extended to complicated models.
\subsection{Owner-dog Dynamic Game}
First we look at a toy example, which is implemented in Python and all derivatives of nonlinear functions are computed via Tensorflow \cite{tensorflow2015-whitepaper}.

\begin{figure}[!t]
  \centering
  \subfigure[DDP method]
    {
      \includegraphics[width=1.5in]{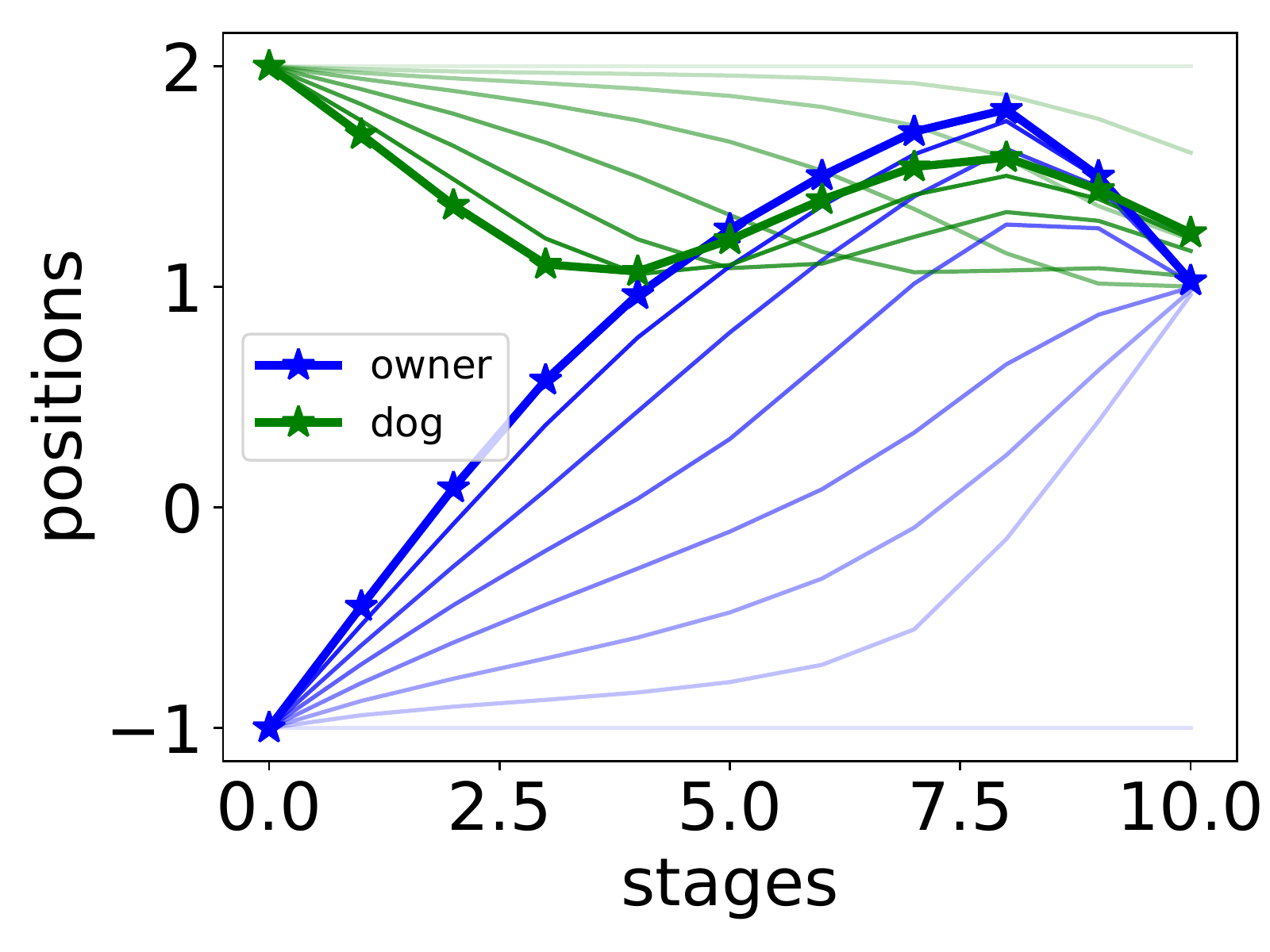}
      \label{fig:od_ddp_traj}
    }
    \subfigure[stagewise Newton step]
    {
      \includegraphics[width=1.5in]{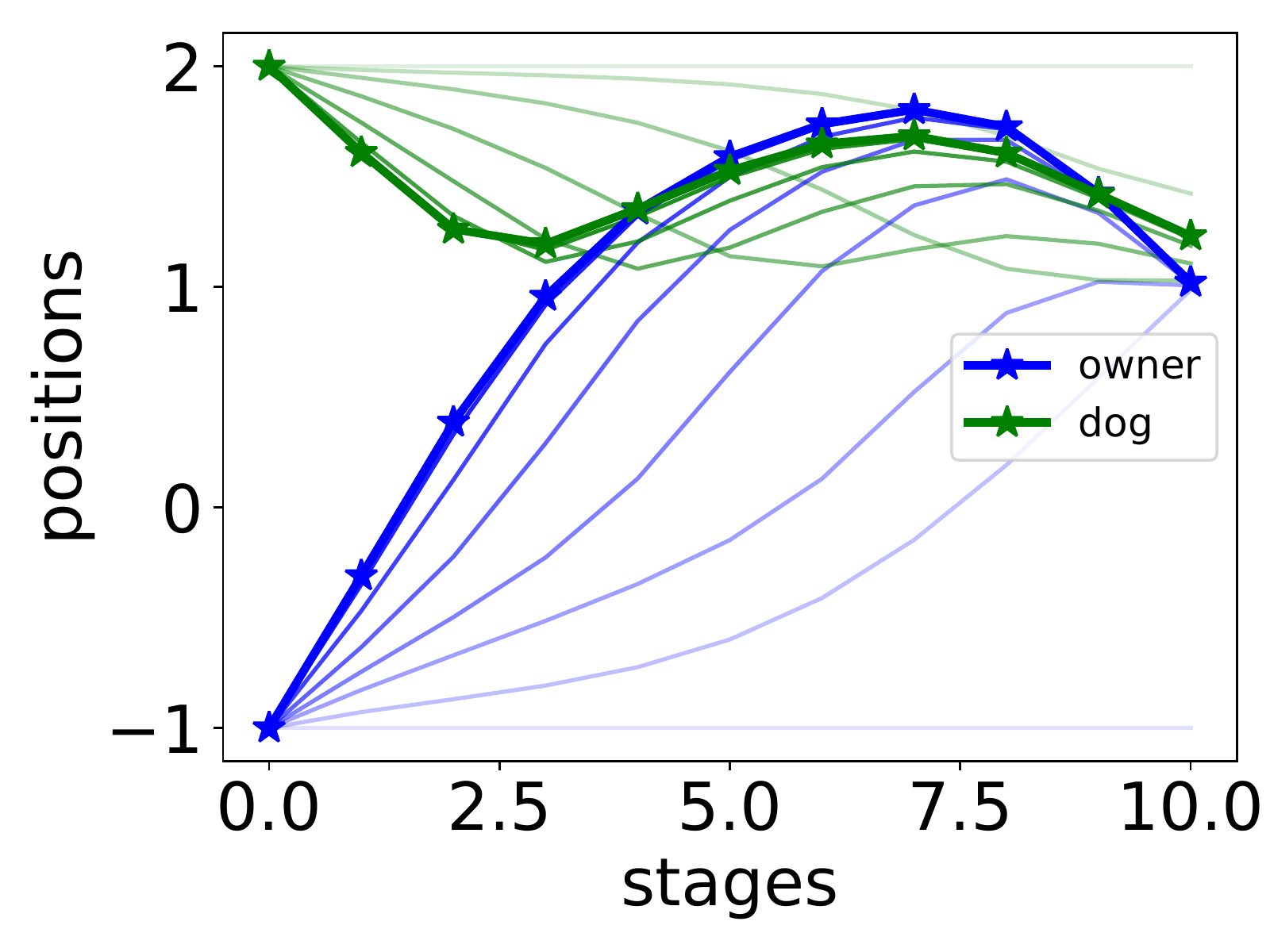}
      \label{fig:od_nm_traj}
    }
  \caption{Owner-dog dynamic game equilibrium trajectories. Lighter colored trajectories are earlier in the overall iterations. The starred trajectory is the final equilibrium solution. We sampled 8 trajectories uniformly spaced out of 300. }
  \label{fig:od}
\end{figure}

\begin{figure}[!t]
  \centering
  \includegraphics[width=2.4in]{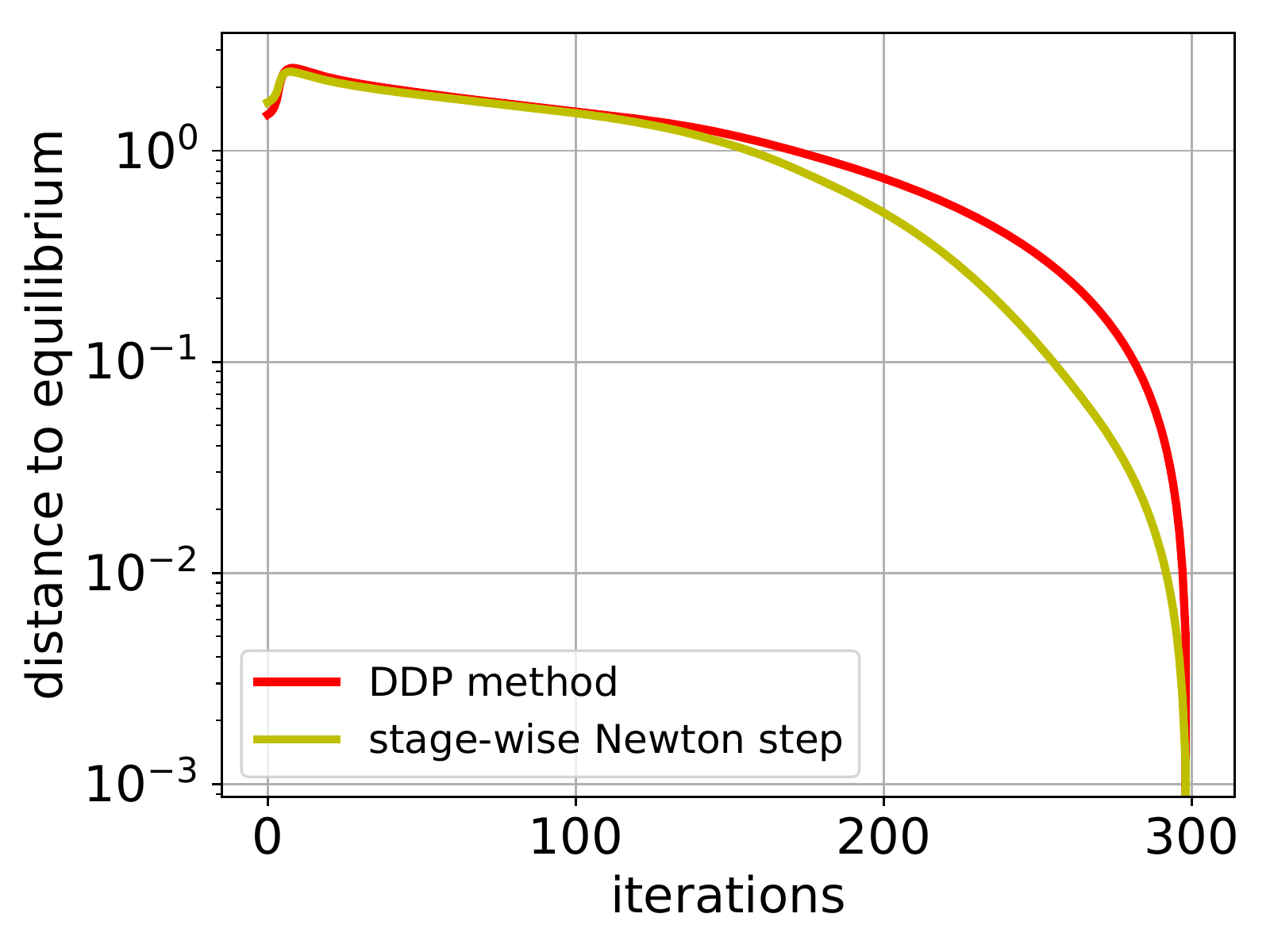}
  \caption{This shows the 2-norm distance between inputs $\bar u$ and the final equilibrium $u^{\star}$ over iterations for both algorithms.}
  \label{fig:od_convergence}
\end{figure}

We consider a simple 1-D owner-dog problem, with horizon $T=10$ and initial state $x_{:,0} = [-1, 2]$ where the dynamics of the owner and the dog are given respectively by
\begin{subequations}
  \begin{align}
    & x_{0, k+1} = x_{0, k} + \tanh (u_{0, k}) \\
    & x_{1, k+1} = x_{1, k} + \tanh (u_{0, k}) \\
    & k = 0, 1, \cdots, T - 1
  \end{align}
\end{subequations}

The owner cares about going to $x_{0, k} = 1$ and that the dog can stay at $x_{1, k} = 2$. The dog, however, only tries to catch up with the owner. Each player also concerns themselves with the energy consumption, therefore has a cost term related to the magnitude of its input. Their cost functions are formulated as
\begin{subequations}
  \begin{align}
    & c_{0,k}(x, u) = \sigmoid ((x_{0, k} - 1)^2) + 40 (x_{1, k} - 2)^2 + (u_{0, k})^2 \\
    & c_{1,k}(x, u) = \tanh^2(x_{0, k} - x_{1, k}) + (u_{1, k})^2 \\
    & k = 0, 1, ..., T - 1
  \end{align}
\end{subequations}
We use a different terminal cost that penalizes much more heavily the owner for not reaching to their target $x_{0, T} = 1.$
\begin{subequations}
  \begin{align}
    & c_{0,T}(x, u) = 100 \ \sigmoid ((x_{0, T} - 1)^2) + 40 (x_{1, T} - 2)^2 \\
    & c_{1,T}(x, u) = \tanh^2(x_{0, T} - x_{1, T})
  \end{align}
\end{subequations}

Nonlinear functions are added to the dynamics and costs to create a nonlinear game rather than for explicit physical meaning. We initialize a trajectory with zero input and initial state, i.e. $\bar u = [0., 0., \ldots, 0.]$ and $\bar x = [-1, 2, -1, 2, \ldots, -1, 2]$. We used an identity regularization matrix with a magnitude of $\lambda = 30$ as in \eqref{eq:regularizeGamma} and performed 300 iterations from the initial trajectory. Note that we started the iteration with a trajectory that is far from a local equilibrium, therefore we do not expect the updates generated by both algorithms to be close.

Fig. \ref{fig:od} shows the solutions found via both algorithms. In order to keep the dog around $x_{1, k} = 2$, the owner has to overshoot and then come back to $x_{0, T}=1$. The dog learns to get closer to the owner over iterations, which is what we would expect given how the problem is formulated. The stagewise Newton's step generated a smoother trajectory in this particular case. Fig. \ref{fig:od_convergence} shows the distances of input to the final equilibrium over all 300 iterations. As can be seen that the error reduces sub-linearly on a log scaled plot, which is evidence that the algorithms converge quadratically. The stagewise Newton's method converges quicker in this particular case. Note that the two methods did not converge to the same trajectory, which is because we fixed a regularization $\lambda$ and started far off the equilibrium. By tuning the regularization with iterations or start from a closer trajectory to the equilibrium, we should improve the situation, which is beyond the scope of this paper.

\subsection{Planar Robots Target Reaching}
Here we consider an experimental setup in which three  planar robots
try to reach each of their own targets, while avoiding collisions with
other robots. The problem is set up such that, if the robots ignore the existence of others and run its own optimal trajectory, they will collide. We apply the proposed DDP algorithm for game to solve for the equilibrium trajectories. This example is implemented in Python and derivatives are computed via PyTorch \cite{paszke2017automatic}.

Robots are modeled as circles on a plane with the location of its
center and a diameter. All robots share the same dynamics given by \eqref{eq:pr_dynamics}. The state $x_{:, k}$ collects all vehicles' positions and $x_{n,k}$ picks the $n$th vehicle's position at step $k$.
\begin{align}
  \label{eq:pr_dynamics}
  x_{n,k+1} = x_{n,k} + (\tanh u_{n,k}) \text{d}t, \quad n = 1, 2, \ldots, N
\end{align}
where $n$ enumerates all robots.
The step cost of each robot consists of a goal cost, a control cost
and, and an avoidance cost. To compute the avoidance cost, we check the distances among robots at each step, when the robots do collide and the distances become
negative, we set it to the small positive number, $0.01$, for
numerical stability. Cost functions follow \eqref{eq:cost_pr}.
\begin{align}
  \label{eq:cost_pr}
  c_{n,k}(x_k, u_k) =& \alpha  (1 - e^{ - ||x_{n,k} - g_{n}||^2}) + ||u_{n,k}||^2 + \nonumber \\
  & \beta \sum_{i \neq n}[ - \log (1 - e^{ - \max(||x_{i,k} - x_{n,k}|| - r_i - r_n, \  0.01)})] \\
  n = & 1, 2, 3
\end{align}
where $g_n$ is the target of the $n$th vehicle, $r_n$ is the radius of the $n$th robot, which are constants given the problem, $\alpha$ and $\beta$ are parameters controlling the relative weights of these three cost terms. The costs are coupled via the distances between robots.

\begin{figure}[!t]
  \centering
  \includegraphics[width=8cm]{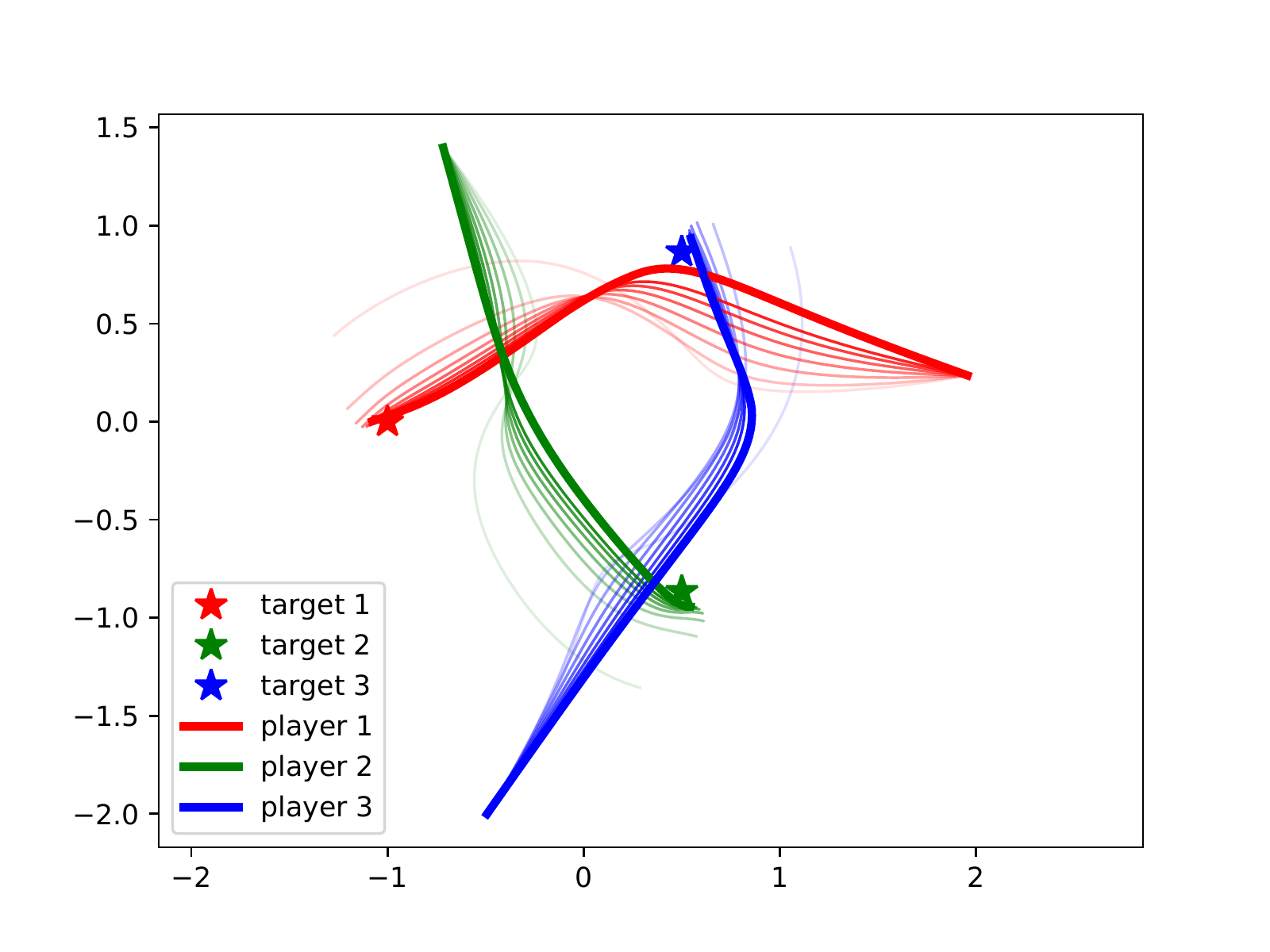}
  \caption{Robots trajectories over iterations. As can be seen that the robots are taking indirect routes to their targets to avoid colliding into each other. As we optimize over the trajectory via the proposed algorithm, the trajectory becomes smoother and the end location closer to the targets. }
  \label{fig:traj_pr}
\end{figure}

\begin{figure}[!t]
  \centering
  \includegraphics[width=8cm]{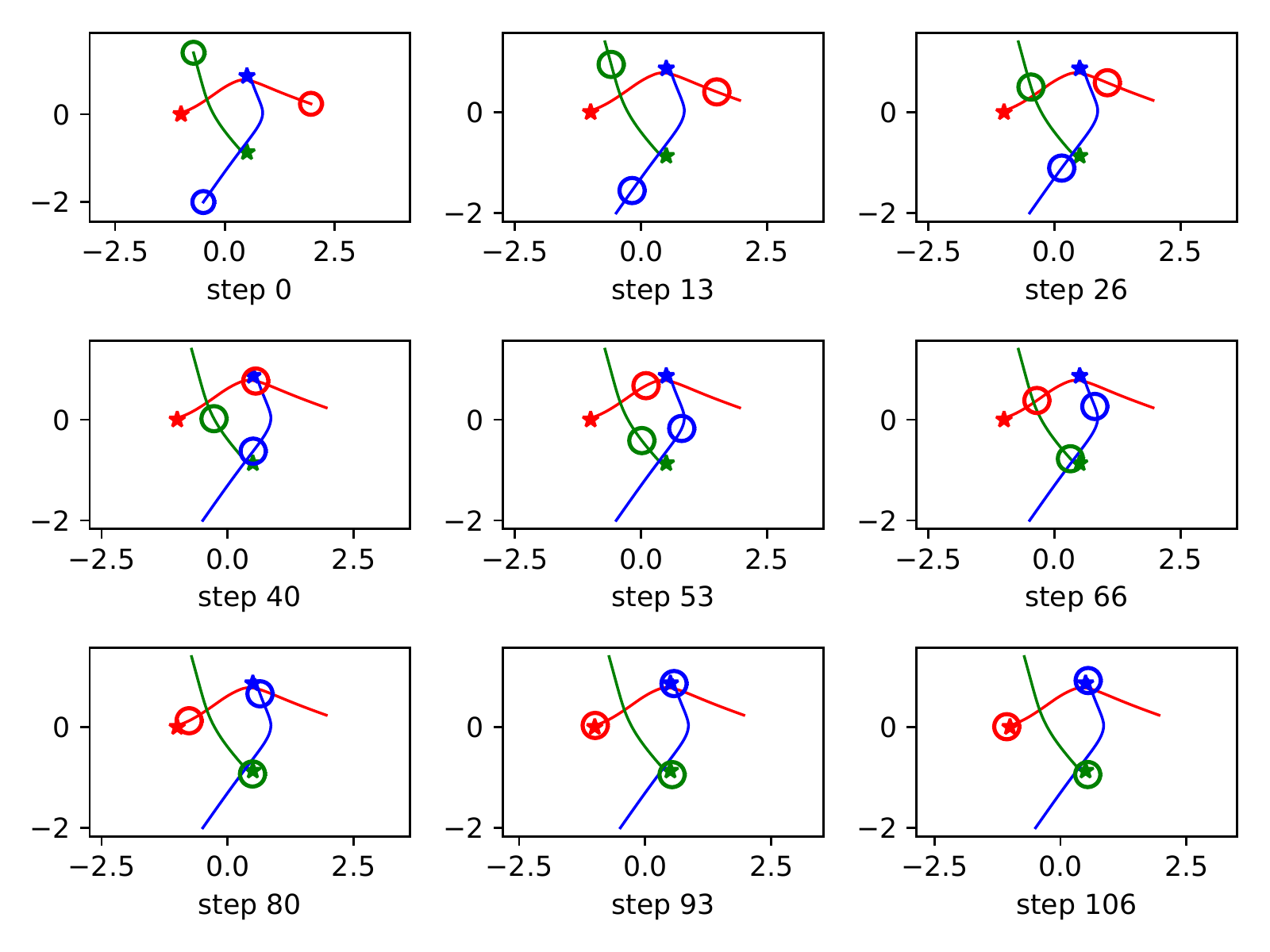}
  \caption{Snapshots of robots position of equilibrium trajactory. Robots are avoiding each other and keeping proper distances from each other.}
  \label{fig:avoiding_sp_pr}
\end{figure}

\begin{figure}[!t]
  \centering
  \includegraphics[width=8cm]{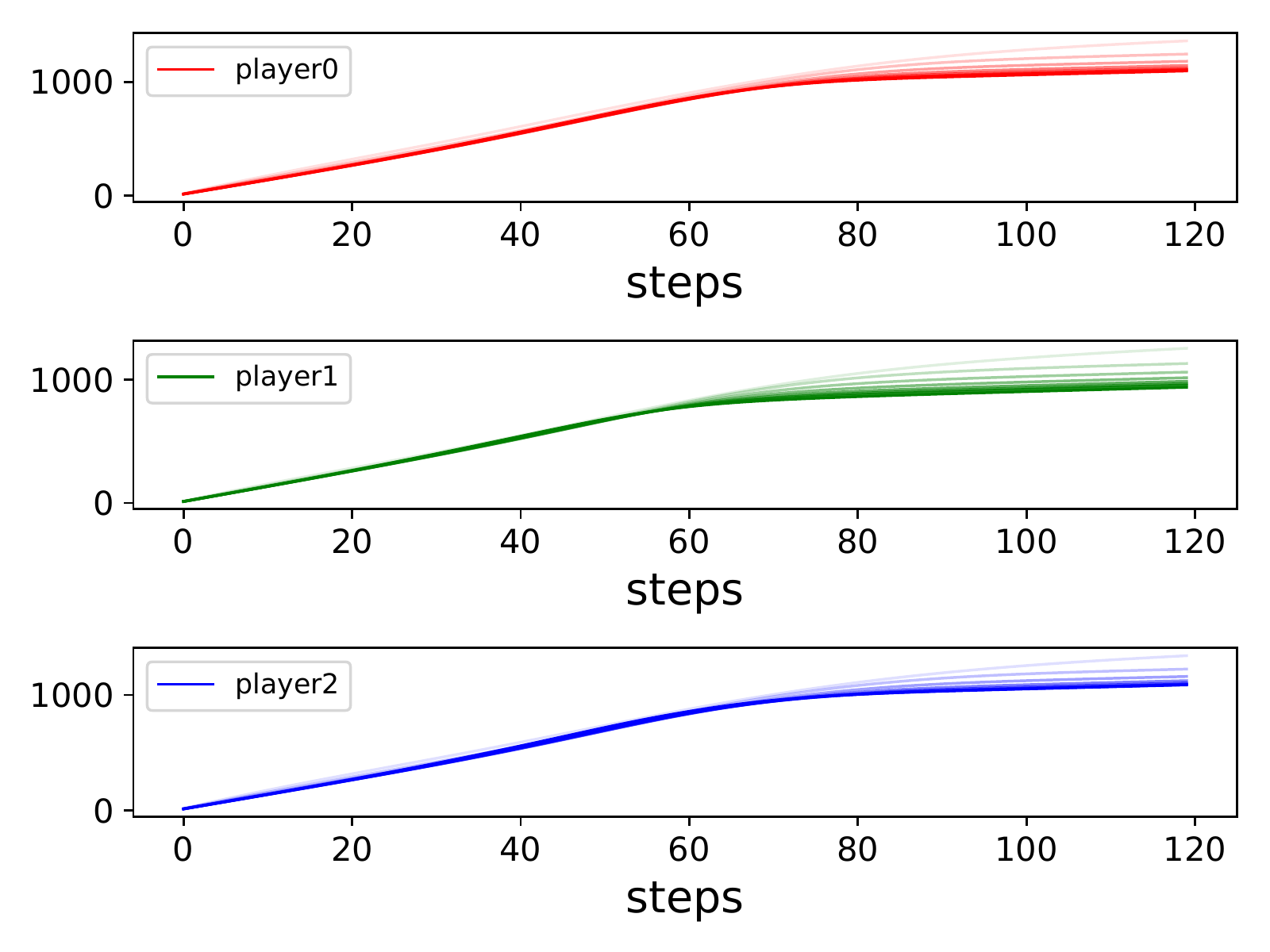}
  \caption{Cumulative costs. The cumulative cost for all robots reduces over iterations. }
  \label{fig:cumsum_cost_pr}
\end{figure}

In our particular implementation, we used the parameters $T = 119$, $\text{d}t =
0.04$, $\alpha = 10$, $\beta = 3$ and $r_i = 0.25$ for all
robots. The targets are chosen as $g_1 = [-1, 0]^\top$, $g_2 = [0.5, -0.866]^\top$ and $g_3 = [0.5, 0.866]^\top$. The robots are initialize at $x_{0} = [1.96, 0.24, -0.72, 1.39, -0.49, -2.00]^\top$.
Figures \ref{fig:traj_pr}, \ref{fig:avoiding_sp_pr} and
\ref{fig:cumsum_cost_pr} show the result of implementing DDP method
to this problem where lighter color means values from earlier
iterations. An initial trajectory was generated with a na\"ive push-pull
control, which assumes each robot is being pulled to its target by an
input that is proportional to the distance to the target, and being
pushed away from other robots by an input that is inversely
proportional to the squared distance to the other robots. This simple control scheme enables the robots to reach their targets given sufficient horizon but requires large inputs and takes sharper turns therefore is far from optimal. Collisions did not happen in any iteration. A regularization magnitude of $\lambda = 10$ as in \eqref{eq:regularizeGamma} was implemented in this example. Due to heavy computational complexity, we only ran the algorithm for 7 iterations. Computing Jacobians in \eqref{eq:Qquad} took up the majority of the program's runtime.

\section{Conclusion and Future Directions}
\label{sec:conclusion}

In this paper we have shown how Newton's method and differential dynamic programming
extends to dynamic games. Convergence of the methods are proven and nature of the equilibria studied. A key step involved was finding explicit
forms for both DDP and stagewise Newton iterations that enable clean comparison
of their solutions. We demonstrated the performance of both algorithms
with nonlinear dynamic games in simulation.
Many extensions are possible. We will examine larger examples and work
on numerical scaling. Derivative-free methods with convergence
guarantee are also attractive since they can be computationally
faster and eliminate the dependence
on analytical models. Both methods can be applied as part of projected gradient descent style methods or operator splitting method \cite{o2013splitting} to constrained dynamic games. Additionally, handling scenarios in which agents have imperfect
model information will be of great practical importance.

\balance

\bibliographystyle{unsrt}
\bibliography{ref.bib}

\balance

\appendix

\section{Auxiliary Proofs}
\subsection{Proof of Theorem \ref{thm:fne}}
\label{app:fne}
We first prove the results corresponding to stagewise Newton method.
Fix a player $n$, and assume that the other players are using the strategy profile $\hat \phi_{-n, t:}$ for a subgame $t$.
Then the optimal policy of player $n$ that minimizes $J_{n, t:}(x_t, \phi_{n, t:},\phi^\star_{-n, t:})$
can be computed from the following optimal control problem:
\begin{subequations}
\label{eq:playerOC}
\begin{align}
  \min_{u_{n}} \quad & \sum_{k=t}^T
                       c_{n,k}(x_k, u_{:,k}) \\
   \textrm{s.t.} \quad & u_{-n,k} = \hat \phi^\star_{-n,k}(x_k) \\
                       & x_{k+1} = f_x(x_k, u_{:,k}) \\
                       & k = t, t+1, ..., T - 1 \\
                       & x_t \textrm{ is given.}
\end{align}
\end{subequations}
The optimal control problem is reduced from the dynamic game by
substituting the equilibrium policy of other players.
To show that the stage-wise Newton method is an approximate feedback
equilibrium, it suffices to show that $\hat \phi_n,t:$ is
approximately optimal for this problem, for all $n$ and $t$.

The quadraticization of problem~\eqref{eq:playerOC} around $[\bar x, \bar u]$ is the same as the
part of player $n$ in the quadraticized dynamic game as in
\eqref{eq:newton_dp}. Since it is assumed that the other players are
playing their equilibrium strategies for the quadraticized game, the
optimal strategy for this quadraticized control problem is precisely
given by player $n$'s solution from the stagewise Newton, $\hat
\phi_{n, t:}$. 
The approximate optimality of $\hat \phi_{n,t:}$  now follows from
Lemma~\ref{lem:opt_fne} below.

DDP is quadratically close to stagewise Newton as evident by the proof of Theorem~\ref{thm:main}, therefore shares the same feedback $O(\epsilon^2)-$Nash equilibrium as stagewise Newton. \hfill\qed

\subsection{Parametric Unconstrained Optimization Lemma}
\label{app:static_FNE}
\begin{lemma}
  \label{lem:opt_fne}
  Given an unconstrained optimization problem with a differentiable objective
  \begin{align}
    \label{eq:arbitrary_opt}
    \min_{x} f(x, p)
  \end{align}
  According to the implicit function theorem, there exists a feedback policy $x = \phi(p)$ that solves the necessary condition of the optimization problem
  \begin{align}
    \frac{\partial f(x, p)}{\partial x}  = 0
  \end{align}
  Furthermore, there exists a feedback policy $\delta x = \bar x + K \delta p$ in the neighborhood where $\epsilon = \norm{\delta p}$ is small that solves
  \begin{align}
    \min_{x} \ \left.\normalfont{\qApprox}(f(x, p))\right|_{\bar x, \bar p}
  \end{align}
  where $\bar x = \phi(\bar p)$ and $\frac{\partial f(\bar x, \bar p)}{\partial x}  = 0$. The feedback policy approximates the true solution locally well in the sense
  \begin{align}
    f(\bar x + K \delta p, \bar p + \delta p) = f(\phi(p), p) + O(\epsilon^2)
  \end{align}
\end{lemma}
\begin{pf*}{Proof.}
  We consider $x$ and $p$ values in the neighborhood of $\bar x$ and $\bar p$, define $x = \bar x + \delta x$, $p = \bar p + \delta p$ and $\norm{\delta p} = \epsilon$. All derivatives in this proof are evaluated at $\bar x$ and $\bar p$.
  Expand $x = \phi(p)$
  \begin{subequations}
    \begin{align}
      \bar x + \delta x &= \phi(\bar p) + \frac{\partial \phi}{\partial p} \delta p + O(\epsilon^2) \\
      \delta x &= \frac{\partial \phi}{\partial p} \delta p + O(\epsilon^2) = O(\epsilon)
    \end{align}
  \end{subequations}

  The quadratic approximation is expanded as
  \begin{align}
    \label{eq:quad_approx}
    & \qApprox(f(x, p))_{\bar x, \bar p} = f(\bar x, \bar p) + \nonumber \\
    & \left.
    \begin{bmatrix}
      \frac{\partial f}{\partial x} & \frac{\partial f}{\partial p}
    \end{bmatrix}\right|_{\bar x, \bar p}
    \begin{bmatrix}
      \delta x \\
      \delta p
    \end{bmatrix} + \frac{1}{2}
    \begin{bmatrix}
      \delta x \\
      \delta p
    \end{bmatrix}^\top \left.
    \begin{bmatrix}
      \frac{\partial^2 f}{\partial x^2} & \frac{\partial^2 f}{\partial x \partial p} \\
      \frac{\partial^2 f}{\partial p \partial x} & \frac{\partial^2 f}{\partial p^2}
    \end{bmatrix} \right|_{\bar x, \bar p}
    \begin{bmatrix}
      \delta x \\
      \delta p
    \end{bmatrix}
  \end{align}
  It can be seen that the $\delta x$ minimizes the approximation is
  \begin{align}
    \delta x = - \left( \frac{\partial^2 f}{\partial x^2} \right)^{-1} \frac{\partial^2 f}{\partial x \partial p} \delta p - \left( \frac{\partial^2 f}{\partial x^2} \right)^{-1} \frac{\partial f}{\partial x} = K \delta p + s
  \end{align}
  where all derivatives are evaluated at $\bar x$, $\bar p$. Because $\frac{\partial f(\bar x, \bar p)}{\partial x}  = 0$, the constant term $s$ is zero. Therefore $\delta x = K \delta p$ minimizes the quadratic approximation.

  Two inequalities come naturally from $\phi(p)$ and $K \delta p$ are minimizers of \eqref{eq:arbitrary_opt}\eqref{eq:quad_approx}.
  \begin{subequations}
    \label{eq:optimizers}
    \begin{align}
      f(\phi(p), p) &\leq f( \bar x + K \delta p, \bar p + \delta p) \\
      \left. \qApprox\left( f( \bar x + K \delta p, \bar p + \delta p) \right)\right|_{\bar x, \bar p} &\leq \qApprox \left. \left( f(\phi(p), p) \right)\right|_{\bar x, \bar p}
    \end{align}
  \end{subequations}
  Taylor series expansion
  \begin{subequations}
    \label{eq:quadf}
    \begin{align}
      f(\phi(p), p) &= f(\phi(\bar p) + \frac{\partial \phi}{\partial p} \delta p + O(\epsilon^2), \bar p + \delta p + O(\epsilon^2)) \\
      &= f(\bar x, \bar p) + \frac{\partial f}{\partial x} \frac{\partial \phi}{\partial p} \delta p + \frac{\partial f}{\partial p} \delta p + O(\epsilon^2) \\
      &= f(\bar x, \bar p) + \frac{\partial f}{\partial p} \delta p + O(\epsilon^2)
    \end{align}
  \end{subequations}
  Compare \eqref{eq:quadf} and \eqref{eq:quad_approx}, and similar comparison can be done for $\left. \qApprox\left(f( \bar x + K \delta p, \bar p + \delta p) \right)\right|_{\bar x, \bar p}$ and $f( \bar x + K \delta p, \bar p + \delta p)$, we get the closeness results
  \begin{subequations}
    \label{eq:value_close}
    \begin{align}
      & \qApprox \left. \left( f(\phi(p), p) \right)\right|_{\bar x, \bar p} = f(\phi(p), p) + O(\epsilon^2) \\
      & \left. \qApprox\left(f( \bar x + K \delta p, \bar p + \delta p) \right)\right|_{\bar x, \bar p} = f( \bar x + K \delta p, \bar p + \delta p) + O(\epsilon^2)
    \end{align}
  \end{subequations}
  Based on \eqref{eq:value_close} and \eqref{eq:optimizers}, we have the following inequalities
  \begin{subequations}
    \begin{align}
      f(\phi(p), p) &\leq f( \bar x + K \delta p, \bar p + \delta p) \\
      &= \left. \qApprox\left(f( \bar x + K \delta p, \bar p + \delta p) \right)\right|_{\bar x, \bar p} + O(\epsilon^2) \\
      &\leq \qApprox \left. \left( f(\phi(p), p) \right)\right|_{\bar x, \bar p} = f(\phi(p), p) + O(\epsilon^2) \\
      &= f(\phi(p), p) + O(\epsilon^2)
    \end{align}
  \end{subequations}
  Therefore, $f( \bar x + K \delta p, \bar p + \delta p) = f(\phi(p), p) + O(\epsilon^2)$.
  \hfill\qed
\end{pf*}

\subsection{Background Results}
\label{app:backgound}
We derive a few results that facilitate the convergence proof.

\begin{lemma}
  Let $F_k$ be defined as in \eqref{eq:newton_dp_invert}.
For $\bar u$ in a neighborhood of $u^\star$, $F_k^{-1}$ exists and
$\rho(F_k^{-1})$ is bounded.
\end{lemma}
\begin{pf*}{Proof.}
First, we bound the spectral radius of $F_k^{-1}$ from above
\begin{align}
  \rho(F_k^{-1}) \leq \hat F
\end{align}
where $\hat F$ is a constant. Consider inverting $\nabla_{u} \pazocal{J}(u)$ in Newton's method by successively eliminating $\delta u_{:,k}$ for $k=T,T-1,\ldots,0$. The
$F_k$ matrices are exactly the matrices which would be inverted when eliminating $\delta u_{:,k}$. Since $\nabla_u \pazocal{J}(u)$ is Lipschitz continuous, its eigenvalues are bounded away from zero in a neighborhood of $u^{\star}$. It follows that the eigenvalues of $F_K$ must also be bounded away from zero and $F_k^{-1}$ is bounded above.
\hfill\qed
\end{pf*}

\begin{lemma}
  \label{lem:omega}
  The following holds
\begin{align}
  \label{eq:omega_equiv}
  \Omega_{n,k} = \frac{\partial}{\partial x_k} \sum_{i=k}^{T} c_{n,i}(x_i, u_{:,i})\Big|_{\bar x, \bar u}
\end{align}
\end{lemma}

\begin{pf*}{Proof.}
$\Omega_{n,k}$ is constructed according to
(\ref{eq:newton_dp_solution_omega1}). Equation (\ref{eq:omega_equiv})
is true for $k = T$ by construction. We proof by induction and assume
that (\ref{eq:omega_equiv}) holds for $k+1$, i.e. $\Omega_{n,k+1} =
\frac{\partial}{\partial x_{k+1}} \sum_{i={k+1}}^{T} c_{n,i}(x_i,
u_{:,i}) \Big|_{\bar x, \bar u}$, then
\begin{subequations}
  \begin{align}
    & \Omega_{n,k} = M_{n,k}^{1x} + \Omega_{n, k+1} A_k \\
    & = \frac{\partial c_{n,k}(x_k, u_{:,k})}{x_k}\Big|_{\bar x, \bar u} + \left( \frac{\partial}{\partial x_{k+1}} \sum_{i=k+1}^{T} c_{n,i}(x_i, u_{:,i}) \right) \frac{\partial x_{k+1}}{\partial x_{k}} \Big|_{\bar x, \bar u} \\
    & = \frac{\partial c_{n,k}(x_k, u_{:,k})}{x_k}\Big|_{\bar x, \bar u} + \left( \frac{\partial}{\partial x_{k}}\Big|_{\bar x, \bar u} \sum_{i=k+1}^{T} c_{n,i}(x_i, u_{:,i}) \right) \\
    & = \frac{\partial}{\partial x_k}\sum_{i=k}^{T} c_{n,i}(x_i,
      u_{:,i}) \Big|_{\bar x, \bar u}
  \end{align}
\end{subequations}
Therefore, (\ref{eq:omega_equiv}) holds for $k$. And by induction, all $k = 0,1, \ldots, T$.
\hfill\qed
\end{pf*}

\begin{lemma}
  \label{lem:gradForm}
  The following is true
\begin{align}
  \label{eq:J_grad_small}
  M^{1u}_{n,k} + \Omega_{n,k+1} B_k = \frac{\partial J_n(u)}{\partial u_{:,k}}\Big|_{\bar u} = O(\epsilon)
\end{align}
\end{lemma}

\begin{pf*}{Proof.}
$\frac{\partial J_n(u)}{\partial u_{:,k}}\Big|_{\bar u} = O(\epsilon)$ is true because $J_n(u)$ is twice differentiable hence Lipschitz, i.e.
\begin{align}
  \Big|\Big| \frac{J_n(u)}{\partial u_{:,k}}\Big|_{\bar u} - \frac{J_n(u)}{\partial u_{:,k}}\Big|_{u^{\star}} \Big|\Big| \leq \text{constant} \cdot \|\bar u - u^{\star}\| = O(\epsilon)
\end{align}
The first equality holds because
\begin{subequations}
  \begin{align}
    \frac{\partial J(u)}{\partial u_{:,k}}\Big|_{\bar u} =& \frac{\partial }{\partial u_{:,k}}\Big|_{\bar x, \bar u} \sum_{i=0}^T c_{n,i}(x_i, u_{:,i}) = \frac{\partial }{\partial u_{:,k}}\Big|_{\bar x, \bar u} \sum_{i=k}^T c_{n,i}(x_i, u_{:,i}) \\
    =& \frac{\partial }{\partial u_{:,k}}\Big|_{\bar x, \bar u} c_{n,k}(x_k, u_{:,k}) + \frac{\partial }{\partial u_{:,k}}\Big|_{\bar x, \bar u} \sum_{i=k+1}^T c_{n,i}(x_i, u_{:,i}) \\
    =& M_{n,k}^{1u} + \frac{\partial }{\partial x_{k+1}} \Big|_{\bar x, \bar u} \sum_{i=k+1}^T c_{n,i}(x_i, u_{:,i}) \frac{\partial x_{k+1}}{\partial u_{:,k}} \Big|_{\bar x, \bar u} \\
    =& M^{1u}_{n,k} + \Omega_{n,k+1} B_k
  \end{align}
\end{subequations}
\hfill\qed
\end{pf*}
These results are used implicitly in the later proofs of lemmas.

  \subsection{Proof of Lemma~\ref{lem:newton_game}} \label{app:newton_game}
First, we prove that the dynamics constraints
(\ref{eq:newton_dp_dynamics0}) and (\ref{eq:newton_dp_dynamics1})
are inductive definitions of \eqref{eq:newton_dp_states0} and \eqref{eq:newton_dp_states1}. Note that $x_0$ is fixed so that \eqref{eq:newton_dp_states0} and \eqref{eq:newton_dp_states1} hold at
$k=0$. Now we handle each of the terms inductively. For $\delta x_{k+1}$, we have
  \begin{align}
    & \delta x_{k+1} = \sum_{i=0}^T \frac{\partial x_{k+1}}{\partial u_{:,i}} \Big|_{\bar x, \bar u} \delta u_{:,i} \nonumber \\
    &= \sum_{i=0}^T \frac{\partial f_k(x_k, u_{:,k})}{\partial u_{:,i}} \Big|_{\bar x, \bar u} \delta u_{:,i} \nonumber \\
    &= \frac{\partial f_k(x_k, u_{:,k})}{\partial x_k} \Big|_{\bar x, \bar u} \sum_{i=0}^T  \frac{\partial x_k} {\partial u_{:,k}} \delta u_{:,i} + \frac{\partial f_k(x_k, u_{:,k})}{\partial u_{:,k}} \Big|_{\bar x, \bar u} \delta u_{:,k} \nonumber \\
    & = A_k \delta x_k + B_k \delta u_{:,k}
  \end{align}
  We used the fact that $\frac{\partial f(x_k, u_{:,k})}{\partial
    u_{:,i}}$ is zero unless $i=k$.

  For $\Delta x_{k+1}$, row $l$ is given by:
  \begin{subequations}
    \begin{align}
      & \Delta x_{k+1}^l = \sum_{i = 0}^{T} \sum_{j = 0}^{T} \delta u_{:,i}^\top {\frac{\partial^2 f^l_k(x_k, u_{:,k})}{\partial u_{:,i} \partial u_{:,j}}} \Big|_{\bar{u}} \delta u_{:,j} \\
      & = \sum_{i = 0}^{T} \sum_{j = 0}^{T} \delta u_{:,i}^\top \left( \frac{\partial^2 f^l_k}{\partial u_{:,i} \partial u_{:,j}} + (\frac{\partial x_k}{\partial u_{:,i}})^\top \frac{\partial^2 f^l_k}{\partial x_k^2} \frac{\partial x_k}{\partial u_{:,j}} \right) \Big|_{\bar{u}} \delta u_{:,j} \nonumber \\
      & + \sum_{i = 0}^{T} \sum_{j = 0}^{T} \delta u_{:,i}^\top \left( (\frac{x_k}{\partial u_{:,i}})^\top \frac{\partial^2 f^l_k}{\partial x_k \partial u_{:,j}} + \frac{\partial^2 f^l_k}{\partial u_{:,i} \partial x_k} \frac{x_k}{\partial u_{:,j}} \right) \Big|_{\bar{u}} \delta u_{:,j} \nonumber \\
      & + \sum_{i = 0}^{T} \sum_{j = 0}^{T} \delta u_{:,i}^\top \left( \sum_{p = 1}^{n_x} \frac{\partial f^l_k}{\partial x^p_k} \frac{\partial^2 x^p_k}{\partial u_{:,i} \partial u_{:,j}} \right) \Big|_{\bar{u}} \delta u_{:,j} \\
      & = \delta u_{:,k}^\top \frac{\partial^2 f^l_k}{\partial u_{:,k}^2} \delta u_{:,k} + \delta x_k^\top \frac{\partial^2 f^l_k}{\partial x_k^2} \delta x_k + \delta x_k^\top \frac{\partial^2 f^l_k}{\partial x_k \partial u_{:,k}} \delta u_{:,k} \nonumber \\
      \label{eq:newton_dp_state1_der0}
      & + \delta u_{:,k}^\top \frac{\partial^2 f^l_k}{\partial u_{:,k} \partial x_k} \delta x_k + \sum_{p=1}^{n_x} \frac{\partial f^l_k}{\partial x^p_k} \sum_{i = 0}^{T} \sum_{j = 0}^{T} \delta u_{:,i}^\top {\frac{\partial^2 x^p_k}{\partial u_{:,i} \partial u_{:,j}}} \delta u_{:,j} \\
      \label{eq:newton_dp_state1_der1}
      & =
        \begin{bmatrix}
          \delta x_k \\
          \delta u_{:,k}
        \end{bmatrix}^\top
      G^l_k
      \begin{bmatrix}
        \delta x_k \\
        \delta u_{:,k}
      \end{bmatrix} + \sum_{p=1}^{n_x} A^{lp}_k \Delta x_k^p
    \end{align}
  \end{subequations}
  To get to each terms in (\ref{eq:newton_dp_state1_der0}), we used the fact that
  \begin{subequations}
    \begin{align}
      & \frac{\partial^2 f^l_k}{\partial u_{:,i} \partial u_{:,j}} = 0, \ \text{for}\  i \neq k \ \text{or}\  j \neq k \\
      & \delta x_k = \sum_{i = 0}^T \frac{\partial x_k}{\partial u_{:,i}} \delta u_{:,i} = \sum_{i = 0}^{k-1} \frac{\partial x_k}{\partial u_{:,i}} \delta u_{:,i}
    \end{align}
  \end{subequations}
  To get to (\ref{eq:newton_dp_state1_der1}), we used the fact
  \begin{subequations}
    \begin{align}
      & \frac{\partial f^l_k}{\partial x^p_k} = A^{lp}_k \\
      & \sum_{i = 0}^{T} \sum_{j = 0}^{T} \delta u_{:,i}^\top {\frac{\partial^2 x^p_k}{\partial u_{:,i} \partial u_{:,j}}} \Big|_{\bar{u}} \delta u_{:,j} = \Delta x_k^p \\
      &
        \begin{bmatrix}
          \frac{\partial^2 f^l_k}{\partial x_k^2} & \frac{\partial^2 f^l_k}{\partial x_k   \partial u_{:,k}} \\
          \frac{\partial^2 f^l_k}{\partial x_k \partial u_{:,k}} & \frac{\partial^2 f^l_k}{\partial u_{:,k}^2}
        \end{bmatrix} \Bigg|_{\bar x, \bar{u}} = G^l_k
    \end{align}
  \end{subequations}
  Both $l$ and $p$ are used to pick out the corresponding element for a vector or matrix. $A^{lp}_k$ means the $l$th row and $p$th column of matrix $A_k$. Equation (\ref{eq:newton_dp_state1_der1}) actually describes each element in (\ref{eq:newton_dp_states1}), so we have proven that both are true.

  Next we prove (\ref{eq:newton_dp_objective}) is the quadratic approximation of $J_n(u)$, i.e.
  \begin{align}
    \label{eq:newton_dp_obj_equiv}
    & \text{quad}(J_n(u))_{\bar u} =
      J_n(\bar u) + \frac{\partial J_n(\bar u)}{\partial u} \delta u + \frac{1}{2} \delta u^\top \frac{\partial^2 J_n(\bar u)}{\partial u^2} \delta u \nonumber \\
    &= \frac{1}{2} \sum_{k=0}^T \left(
      \begin{bmatrix}
        1 \\
        \delta x_k \\
        \delta u_{:,k}
      \end{bmatrix}^\top
    M_{n, k}
    \begin{bmatrix}
      1 \\
      \delta x_k \\
      \delta u_{:,k}
    \end{bmatrix}
    + M^{1k}_{n,k} \Delta x_k \right)
  \end{align}
  We need the explicit expressions for the associated derivatives.
  \begin{subequations}
    \label{eq:Jderivatives}
    \begin{align}
      & \frac{\partial J_n(u)}{\partial u_{:,i}} = \sum_{k=0}^T \left( \frac{\partial c_{n, k}(x_k,u_{:,k})}{\partial u_{:,i}} + \frac{\partial c_{n, k}(x_k,u_{:,k})}{\partial x_k} \frac{\partial x_k}{\partial u_{:,i}} \right) \\
      & \frac{\partial^2 J_n(u)}{\partial u_{:,i} \partial u_{:,j}} \nonumber \\
      & = \sum_{k=0}^T \left( \frac{\partial^2 c_{n, k}(x_k,u_{:,k})}{\partial u_{:,i} \partial u_{:,j}} + \frac{\partial x_k}{\partial u_{:,i}}^\top \frac{\partial^2 c_{n, k}(x_k,u_{:,k})}{\partial x_k^2} \frac{\partial x_k}{\partial u_{:,j}} \right) \nonumber \\
      & + \sum_{k=0}^T \left( \frac{\partial x_k}{\partial u_{:,i}}^\top \frac{\partial^2 c_{n, k}(x_k,u_{:,k})}{\partial x_k \partial u_{:,j}} + \frac{\partial^2 c_{n, k}(x_k,u_{:,k})}{\partial u_{:,i} \partial x_k} \frac{\partial x_k}{\partial u_{:,j}} \right) \nonumber \\
      \label{eq:smallHessian}
      & +\sum_{k=0}^T \sum_{l=1}^{n_x} \frac{\partial c_{n, k}(x_k,u_{:,k})}{\partial x_k^l} \frac{\partial^2 x_k^l}{\partial u_{:,i} \partial u_{:,i}}
    \end{align}
  \end{subequations}

  We break down each term in (\ref{eq:newton_dp_obj_equiv}). First the second order term.
  \begin{subequations}
    \label{eq:newton_dp_objective_second}
    \begin{align}
      & \delta u^\top \frac{\partial^2 J_n(\bar u)}{\partial u^2} \delta u =
        \sum_{i,j=0}^T \delta u_{:,i}^T \frac{\partial^2 J_n(u)}{\partial u_{:,i} \partial u_{:,j}} \bigg|_{\bar u} \delta u_{:,j} \\
      & =  \sum_{i,j,k=0}^T \delta u_{:,i}^\top \left(
        \frac{\partial^2 c_{n,k}(x_k,u_{:,k})}{\partial u_{:,i} \partial u_{:,j}} \bigg|_{\bar u} \right) \delta u_{:,j} \nonumber \\
      & + \sum_{i,j,k=0}^T \delta u_{:,i}^\top \left( \frac{\partial x_k}{\partial u_{:,i}}^\top
        \frac{\partial^2 c_{n,k}(x_k,u_{:,k})}{\partial x_k^2} \bigg|_{\bar u}
        \frac{\partial x_k}{\partial u_{:,j}}
        \right) \delta u_{:,j} \nonumber \\
      & + \sum_{i,j,k=0}^T \delta u_{:,i}^\top \left(
        \frac{\partial x_k}{\partial u_{:,i}}^\top
        \frac{\partial^2 c_{n,k}(x_k,u_{:,k})}{\partial x_k \partial u_{:,j}} \bigg|_{\bar u} \right) \delta u_{:,j} \nonumber \\
      \label{eq:newton_dp_objective_sec0}
      & + \sum_{i,j,k=0}^T \delta u_{:,i}^\top \left( \frac{\partial^2 c_{n,k}(x_k,u_{:,k})}{\partial u_{:,i} \partial x_k} \bigg|_{\bar u}
        \frac{\partial x_k}{\partial u_{:,j}} \right) \delta u_{:,j} \nonumber \\
      & + \sum_{k=0}^T \sum_{p=1}^n
        \frac{\partial c_{n,k}(x_k,u_{:,k})}{\partial x_k^p} \bigg|_{\bar u}
        \sum_{i,j=0}^T \delta u_{:,i}^\top \frac{\partial^2 x_k^p}{\partial u_{:,i} \partial u_{:,i}} \delta u_{:,j} \\
      \label{eq:newton_dp_objective_sec2}
      = & \sum_{k=0}^T \left( \delta u_{:,k}^\top \frac{\partial^2 c_{n,k}}{\partial u_{:,k}^2 } \bigg|_{\bar u} \delta u_{:,k} + \delta x_k^\top \frac{\partial^2 c_{n,k}}{\partial x_k^2} \bigg|_{\bar u} \delta x_k \right) + \nonumber \\
      & \sum_{k=0}^T  \left( \delta x_k ^\top \frac{\partial^2 c_{n,k}}{\partial x_k \partial u_{:,k}}\vert_{\bar u}\delta u_{:,k} + \delta u_{:,k} ^\top \frac{\partial^2 c_{n,k}}{\partial u_{:,k} \partial x_k} \bigg|_{\bar u} \delta x_k\right) + \nonumber \\
      & \sum_{k=0}^T \sum_{p=1}^n
        \frac{\partial c_{n,k}(x_k,u_{:,k})}{\partial x_k^p} \Delta x_k^p \\
      = & \sum_{k=0}^T \left(
          \begin{bmatrix}
            \delta x_k \\
            \delta u_{:,k}
          \end{bmatrix}^\top
      \begin{bmatrix}
        \frac{\partial^2 c_{n,k}}{\partial x_k^2}
        & \frac{\partial^2 c_{n,k}}{\partial x_k \partial u_{:,k}} \\
        \frac{\partial^2 c_{n,k}}{\partial u_{:,k}\partial x_k}
        & \frac{\partial^2 c_{n,k}}{\partial u_{:,k}^2}
      \end{bmatrix} \bigg|_{\bar u}
          \begin{bmatrix}
            \delta x_k \\
            \delta u_{:,k}
          \end{bmatrix} \right) \nonumber \\
      & + \left( \frac{\partial c_{n,k}(x_k,u_{:,k})}{\partial x_k} \Delta x_k \right)
    \end{align}
  \end{subequations}
  The first term in \eqref{eq:newton_dp_objective_sec0} to \eqref{eq:newton_dp_objective_sec2} holds because $c_{n,k}(x_k,u_{:,k})$ only depends directly on $u_{:,i}$ and $u_{:,j}$ when $i=j=k$. The others hold because  $x_k$ only depends on $u_{:,i}$ and $u_{:,j}$ when $i,j <k$. The last term uses the definition of $\Delta x_k$ in \eqref{eq:newton_dp_states1}.

  The first order term
  \begin{subequations}
    \label{eq:newton_dp_objective_first}
    \begin{align}
      & \frac{\partial J_n(\bar u)}{\partial u} \delta u = \sum_{i=0}^T \frac{\partial J_n(u)}{\partial u_{:,i}} \delta u_{:,i} \\
      & = \sum_{i,k=0}^T \left( \frac{\partial c_{n,k}(x_k,u_{:,k})}{\partial u_{:,i}} + \frac{\partial c_{n,k}(x_k,u_{:,k})}{\partial x_k} \frac{\partial x_k}{\partial u_{:,i}} \right) \delta u_{:,i} \\
      & = \sum_{k=0}^T \left( \frac{\partial c_{n,k}(x_k,u_{:,k})}{\partial u_{:,k}} \delta u_{:,k} + \frac{\partial c_{n,k}(x_k,u_{:,k})}{\partial x_k} \sum_{i=0}^T \frac{\partial x_k}{\partial u_{:,i}} \right) \\
      & = \sum_{k=0}^T \left( \frac{\partial c_{n,k}(x_k,u_{:,k})}{\partial u_{:,k}} \delta u_{:,k} + \frac{\partial c_{n,k}(x_k,u_{:,k})}{\partial x_k} \delta x_k\right)
    \end{align}
  \end{subequations}

  And constant term
  \begin{align}
    \label{eq:newton_dp_objective_constant}
    J_n(\bar u) = \sum_{k=0}^T c_{n,k}(\bar x_k, \bar u_{:,k})
  \end{align}
  From (\ref{eq:newton_dp_objective_second}),
  (\ref{eq:newton_dp_objective_first}), and
  (\ref{eq:newton_dp_objective_constant}) it follows that (\ref{eq:newton_dp_obj_equiv}) is true.
\hfill\qed

\subsection{Proof of Lemma~\ref{lem:newtonMatrices}}
\label{app:newtonMatrices}
Substituting the approximated dynamic game \eqref{eq:newton_dp} to the Bellman equation \eqref{eq:bellman} leads us to,
\begin{subequations}
  \label{eq:newtonRecursion}
  \begin{align}
    \hat V^{\bar u}_{n,T+1}(\delta x_{T+1},\Delta x_{T+1}) &= 0 \\
    \nonumber
    \hat Q^{\bar u}_{n,k}(\delta x_k,\Delta x_k,\delta u_{:,k})
                                             &=
    \\
                                             &
                                               \hspace{-8em}
                                               \frac{1}{2}
                                               \left(
                                               \begin{bmatrix}
                                                 1 \\
                                                 \delta x_k \\
                                                 \delta u_{:,k}
                                               \end{bmatrix}^\top
    M_{n,k}
    \begin{bmatrix}
      1 \\
      \delta x_k \\
      \delta u_{:,k}
    \end{bmatrix}
    +M^{1x}_{n,k} \Delta x_k \right) \nonumber \\
                                             &
                                               \hspace{-8em}
                                               + \hat V^{\bar u}_{n,k+1}(A_k\delta x_k +B_k
                                               \delta u_{:,k},
                                               A_k \Delta x_k + R_k(\delta x_k,\delta u_{:,k})) \\
    \label{eq:newtonQgame}
    \hat V^{\bar u}_{n,k}(\delta x_k,\Delta x_k) &= \min_{
                                     \delta u_{n,k}} \hat Q^{\bar u}_{n,k}(\delta
                                     x_k,\Delta x_k,\delta u_{:,k}).
  \end{align}
\end{subequations}
Note that \eqref{eq:newtonQgame} defines a static quadratic game and $\hat V^{\bar u}_{n,k}(\delta x_k, \Delta x_k)$ is found by solving the game and substituting the solution back to $\hat Q^{\bar u}_{n,k}(\delta x_k, \Delta x_k, \delta u_{:,k})$.

Solving the equilibrium strategy and $\hat V^{\bar u}_{n,k}(\delta x_k, \Delta x_k)$ based on $\hat Q^{\bar u}_{n,k}(\delta x_k, \Delta x_k, \delta u_{:,k})$ is the same as how we arrived at \eqref{eq:ddp_necessary} and \eqref{eq:TildeSDef}, since the extra terms of $\Delta x_k$ are not coupled with $\delta u_{:,k}$ and other terms are of the exact same form.
The stepping back in time of $\hat Q^{\bar u}_k(\delta x_k, \Delta x_k, \delta u_{:,k})$ is acheived by substituting \eqref{eq:newton_dp_dynamics0} and \eqref{eq:newton_dp_dynamics1} into \eqref{eq:newton_dp_solution_state_val_fun}, which is slightly different because of the extra terms related to $\Delta x_k$.
\begin{subequations}
  \label{eq:newton_dp_backprop}
  \begin{align}
    & \hat Q^{\bar u}_{n,k}(\delta x_k, \Delta x_k, \delta u_{:,k}) \\
    &= \frac{1}{2} \left(
      \begin{bmatrix}
        1 \\
        \delta x_k \\
        \delta u_{:,k}
      \end{bmatrix}^\top
    M_{n,k}
    \begin{bmatrix}
      1 \\
      \delta x_k \\
      \delta u_{:,k}
    \end{bmatrix} + M^{1k}_{n,k} \Delta x_k \right) \nonumber \\
    & \quad + \hat V^{\bar u}_{n,k+1}(A_k \delta{x}_k + B_k  \delta{u}_k, A_k \Delta x_{k}  + R_k(\delta x_k, \delta u_{:,k})) \\
    &= \frac{1}{2} \left(
      \begin{bmatrix}
        1 \\
        \delta x_k \\
        \delta u_{:,k}
      \end{bmatrix}^\top
    M_{n,k}
    \begin{bmatrix}
      1 \\
      \delta x_k \\
      \delta u_{:,k}
    \end{bmatrix} \right) + \nonumber \\
    & \frac{1}{2}
      \begin{bmatrix}
        1 \\
        \delta{x}_k \\
        \delta u_{:,k}
      \end{bmatrix}^\top
    \begin{bmatrix}
      S^{11}_{n,k+1} & S^{1x}_{n,k+1} A_k & S^{1x}_{n,k+1} B_k \\
      A^\top_k S^{x1}_{n,k+1} & A^\top_k S^{xx}_{n,k+1} A_k & A^\top_k S^{xx}_{n,k+1} B_k \\
      B^\top_k S^{x1}_{n,k+1} & B^\top_k S^{xx}_{n,k+1} A_k & B^\top_k S^{xx}_{n,k+1} B_k
    \end{bmatrix}
                                                              \begin{bmatrix}
                                                                1 \\
                                                                \delta{x}_k \\
                                                                \delta u_{:,k}
                                                              \end{bmatrix} \nonumber \\
    & \quad + \frac{1}{2} \left( (M^{1k}_{n,k} + \Omega_{n,k+1} A_k) \Delta x_{k} \right) \nonumber \\
    & \quad + \frac{1}{2} \left(
      \begin{bmatrix}
        \delta x_k \\
        \delta u_{:,k}
      \end{bmatrix}^\top
    D_{n,k}
    \begin{bmatrix}
      \delta x_k \\
      \delta u_{:,k}
    \end{bmatrix} \right)
  \end{align}
\end{subequations}
So (\ref{eq:newton_dp_solution_omega1}),
(\ref{eq:newton_dp_solution_D}) and (\ref{eq:GammaDef}) are true.
\hfill\qed

\subsection{Proof of Lemma~\ref{lem:nmSol}}
\label{app:nmSol}
As discussed in the proof of Lemma~\ref{lem:nmMatrices}, a necessary
condition for the solution of \eqref{eq:newton_dp_solution_state_action_val_fun} is given by
\eqref{eq:nm_necessary}. Thus, a sufficient condition for a unique
solution is that $F_k$ be invertible. At the beginning of the
appendix, we showed that $F_k^{-1}$ exists near $u^\star$ and that
its spectral radius is bounded. So the game definded by \eqref{eq:newton_dp_solution_state_action_val_fun} has a unique solution. \hfill\qed

\subsection{Proof of Lemma~\ref{lem:quadSol}}
\label{app:ddpSol}
As discussed in the proof of Lemma~\ref{lem:nmMatrices}, a necessary
condition for the solution of \eqref{eq:quadQGame} is given by
\eqref{eq:localSolution}. Thus, a sufficient condition for a unique
solution is that $\tilde F_k$ be invertible. At the beginning of the
appendix, we showed that $F_k^{-1}$ exists near $u^\star$ and that
its spectral radius is bounded.

Now we show that $\tilde F_k$ exists and $\rho(\tilde F_k^{-1})$ is bounded.
Lemma~\ref{lem:matrixClose} implies that $\tilde F_k = F_k +
O(\epsilon)$. It follows that
\begin{equation*}
  \tilde F_k^{-1} = (F_k + O(\epsilon))^{-1} = F_{k}^{-1} - F_k^{-1}
  O(\epsilon) F_k^{-1} = F_k^{-1} + O(\epsilon).
\end{equation*}
It follows that $\tilde F_k^{-1}$ exists and is bounded in a
neighborhood of $u^\star$. \qed

\subsection{Closeness Lemmas}
\label{app:closeness_lems}
This section contains lemmas that prove the updates generated by stagewise Newton's method and DDP are close. The first lemma shows that the matrices used in the backwards
recursions are close.

\begin{lemma}
  \label{lem:matrixClose}
  {\it
    The matrices from the backwards recursions of DDP and stagewise Newton's method are close in the following sense:
    \begin{subequations}
      \label{eq:closeness0}
      \begin{align}
        \label{eq:close_D}
        \tilde D_{n,k} &= D_{n,k} + O(\epsilon) \\
        \label{eq:close_S0}
        S^{1x}_{n,k} &= \Omega_{n,k} + O(\epsilon) \\
        \label{eq:close_S1}
        \tilde S^{1x}_{n,k} &= \Omega_{n,k} + O(\epsilon) \\
        \label{eq:close_S2}
        \tilde S^{1x}_{n,k} &= S^{1x}_{n,k} + O(\epsilon^2) \\
        \label{eq:close_S4}
        \tilde S^{11}_{n,k} &= S^{11}_{n,k} + O(\epsilon^2) \\
        \label{eq:close_S3}
        \tilde S_{n,k} &= S_{n,k} + O(\epsilon) \\
        \label{eq:close_gamma0}
        \begin{bmatrix}
          \tilde \Gamma^{1x}_{n,k} & \tilde \Gamma^{1u}_{n,k}
        \end{bmatrix} &=
        \begin{bmatrix}
          \Gamma^{1x}_{n,k} & \Gamma^{1u}_{n,k}
        \end{bmatrix} + O(\epsilon^2) \\
        \label{eq:close_gamma4}
        \tilde \Gamma^{11}_{n,k} &= \Gamma^{11}_{n,k} + O(\epsilon^2) \\
        \label{eq:close_gamma1}
        \tilde \Gamma_{n,k} &= \Gamma_{n,k} + O(\epsilon)\\
        \label{eq:close_gamma2}
        \Gamma_{n,k}^{1u} &= O(\epsilon) \\
        \label{eq:close_gamma3}
        \tilde \Gamma_{n,k}^{1u} &= O(\epsilon)
      \end{align}
    \end{subequations}
    and that
    \begin{subequations}
      \label{eq:closeness1}
      \begin{align}
        \label{eq:close_F}
        \tilde F_k &= F_k + O(\epsilon) \\
        \label{eq:close_P}
        \tilde P_k &= P_k + O(\epsilon) \\
        \label{eq:close_H0}
        \tilde H_k &= H_k + O(\epsilon^2) \\
        \label{eq:close_H1}
        \tilde H_k &= O(\epsilon) \\
        \label{eq:close_H2}
        H_k &= O(\epsilon) \\
        \label{eq:close_s0}
        \tilde s_k &= s_k + O(\epsilon^2) \\
        \label{eq:close_s1}
        \tilde s_k &= O(\epsilon) \\
        \label{eq:close_s2}
        s_k &= O(\epsilon) \\
        \label{eq:close_K}
        \tilde K_k &= K_k + O(\epsilon),
      \end{align}
    \end{subequations}
  }
\end{lemma}

\begin{pf*}{Proof.}
  We give a proof by induction. For $k = T$, because of the way these variables are constructed, they are identical, i.e.
  \begin{subequations}
    \begin{align}
      \Gamma_{n,T} =& \tilde \Gamma_{n,T} \\
      F_T =& \tilde F_T \\
      P_T =& \tilde P_T \\
      H_T =& \tilde H_T \\
      s_T =& \tilde s_T \\
      K_T =& \tilde K_T \\
      S_{n,T} =& \tilde S_{n,T}
    \end{align}
  \end{subequations}
  so we have  (\ref{eq:close_S2}) (\ref{eq:close_S3}) (\ref{eq:close_gamma0}) (\ref{eq:close_gamma1}) (\ref{eq:close_gamma4}) (\ref{eq:close_F}) (\ref{eq:close_P}) (\ref{eq:close_H0}) (\ref{eq:close_s0}) (\ref{eq:close_K}) hold for $k = T$. \\
  We also know that
  \begin{align}
    M_{n,T}^{1u} = \frac{\partial c_{n,T}}{\partial u_T} = \frac{\partial J_n(u)}{\partial u_T}=O(\epsilon)
  \end{align} where the first equality is by construction, the second is true because $u_T$ only appears in $J_n(u)$ in $c_{n,T}$. By construction, $\Gamma_{n,T}^{1u} = \tilde \Gamma_{n,T}^{1u} = M_{n,T}^{1u} = O(\epsilon)$, so (\ref{eq:close_gamma2})(\ref{eq:close_gamma3}) are true for $k = T$.
  Similarly, $H_T$ and $\tilde H_T$ are constructed from $\Gamma_{n,T}^{u1}$ and $\tilde \Gamma_{n,T}^{u1}$, so (\ref{eq:close_H1})(\ref{eq:close_H2}) are true for $k=T$.

  Because $F_k^{-1}$ is bounded above, $s_T = - F_T^{-1} H_T = - F_T^{-1} O(\epsilon) = O(\epsilon)$. Similarly, $\tilde s_T = O(\epsilon)$.
  Equations (\ref{eq:close_s1})(\ref{eq:close_s2}) are true for $k=T$.

  From (\ref{eq:newton_dp_solution_S}) and $\Gamma_{n,T} = M_{n,T}$, we can get
  \begin{subequations}
    \begin{align}
      S_{n,T}^{1x} =& M_{n,T}^{1x} + M_{n,T}^{1u} K_T + s_T^\top ( M_{n,T}^{ux} + M_{n,T}^{uu} K_T ) \\
      =& \Omega_{n,T} + O(\epsilon) K_T + O(\epsilon) ( M_{n,T}^{ux} + M_{n,T}^{uu} K_T ) \\
      =& \Omega_{n,T} + O(\epsilon)
    \end{align}
  \end{subequations}
  because $M_{n,T}$ is bounded. Hence (\ref{eq:close_S0}) is true. Further, (\ref{eq:close_S1}) is also true.

  The time indices for $D_{n,T-1}$ and $\tilde D_{n,T-1}$ go to a
  maximum of $T-1$, so to prove things inductively, we need
  (\ref{eq:close_D}) to hold for $k=T-1$. The difference between
  constructions of $D_{n,T-1}$ and $\tilde D_{n,T-1}$ is in that the
  former uses $\Omega_{n,T}$ and the later uses $\tilde
  S_{n,T}^{1x}$. But since we have proved that $\Omega_{n,T} = \tilde S_{n,T}^{1x} + O(\epsilon)$, and $G_{T-1}$ is bounded, we can also conclude $D_{n,T-1} = \tilde D_{n,T-1} + O(\epsilon)$. Therefore (\ref{eq:close_D}) is true for $k = T-1$.

  So far, we have proved that for the last step, either $k = T$ or $k= T -1$, (\ref{eq:closeness0})\eqref{eq:closeness1} are true. Assuming except for (\ref{eq:close_D}), (\ref{eq:closeness0})\eqref{eq:closeness1} are true for $k+1$ and (\ref{eq:close_D}) is true for $k$. If we can prove all equations hold one step back, our proof by induction would be done.

  Assume (\ref{eq:close_D}) holds for $k$ and other equations in (\ref{eq:closeness0})\eqref{eq:closeness1} hold for $k+1$. Readers be aware that we use these assumptions implicitly in the derivations following.

  From (\ref{eq:GammaDef}) we can get
  \begin{subequations}
    \begin{align}
      \Gamma_{n,k}^{1u} =& M_{k,T}^{1u} + S_{n,k+1}^{1x} B_k \\
      =& M_{k,T}^{1u} + \Omega_{n,k+1} B_k + O(\epsilon) B_k \\
      =& O(\epsilon)
    \end{align}
  \end{subequations}
  Here we used (\ref{eq:J_grad_small}). Similarly, we can prove $\tilde \Gamma_{n,k}^{1u} = O(\epsilon)$. So (\ref{eq:close_gamma2}) and (\ref{eq:close_gamma3}) hold for $k$.

  From (\ref{eq:GammaDef}) and (\ref{eq:TildeGammaBackprop}) we can compute the difference between $\tilde \Gamma_{n,k}$ and $\Gamma_{n,k}$ as
  \begin{subequations}
    \begin{align}
      & \ \tilde \Gamma_{n,k} - \Gamma_{n,k} \nonumber \\
      = &
          \begin{bmatrix}
            \tilde S_{n, k+1}^{11} - S_{n, k+1}^{11} & \mathbf{0} & \mathbf{0} \\
            A_k^\top (\tilde S_{n,k+1}^{x1} - S_{n,k+1}^{x1}) & \mathbf{0} & \mathbf{0} \\
            B_k^\top  (\tilde S_{n,k+1}^{x1} - S_{n,k+1}^{x1}) & \mathbf{0} & \mathbf{0}
          \end{bmatrix} \nonumber \\
      & +
        \begin{bmatrix}
          \mathbf{0} & (\tilde S_{n,k+1}^{1x} - S_{n,k+1}^{1x})A_k & \mathbf{0} \\
          \mathbf{0} & A_k^\top (\tilde S_{n,k+1}^{xx} - S_{n,k+1}^{xx})A_k + (\tilde D_k^{xx} - D_k^{xx}) & \mathbf{0} \\
          \mathbf{0} & B_k^\top (\tilde S_{n,k+1}^{xx} - S_{n,k+1}^{xx}) A_k + (\tilde D_k^{ux} - D_k^{ux}) & \mathbf{0}
        \end{bmatrix} \nonumber \\
      & +
        \begin{bmatrix}
          \mathbf{0} & \mathbf{0} & (\tilde S_{n,k+1}^{1x} - S_{n,k+1}^{1x})B_k \\
          \mathbf{0} & \mathbf{0} & A_k^\top (\tilde S_{n,k+1}^{xx} - S_{n,k+1}^{xx}) B_k + (\tilde D_k^{xu} - D_k^{xu}) \\
          \mathbf{0} & \mathbf{0} & B_k^\top (\tilde S_{n,k+1}^{xx} - S_{n,k+1}^{xx}) B_k + (\tilde D_k^{uu} - D_k^{uu})
        \end{bmatrix} \\
      =& \begin{bmatrix}
        O(\epsilon^2) & A_k O(\epsilon^2) & B_k O(\epsilon^2) \\
        A_k^\top O(\epsilon^2) & A_k^\top O(\epsilon) A_k  + O(\epsilon) & A_k^\top O(\epsilon) B_k  + O(\epsilon) \\
        B_k^\top O(\epsilon^2)  & B_k^\top O(\epsilon) A_k + O(\epsilon) & B_k^\top O(\epsilon) B_k + O(\epsilon)
      \end{bmatrix} \\
      =&
         \begin{bmatrix}
           O(\epsilon^2) & O(\epsilon^2) & O(\epsilon^2) \\
           O(\epsilon^2) & O(\epsilon) & O(\epsilon) \\
           O(\epsilon^2) & O(\epsilon) & O(\epsilon)
         \end{bmatrix}
    \end{align}
  \end{subequations}
  from which we can see that \eqref{eq:close_gamma4}(\ref{eq:close_gamma0}) and (\ref{eq:close_gamma1}) are true.
  Once we proved the closeness between $\tilde \Gamma_{n,k}$ and $\Gamma_{n,k}$ and the specific terms are $O(\epsilon)$, i.e. (\ref{eq:close_gamma0}) to (\ref{eq:close_gamma3}), because of they way they are constructed from $\tilde \Gamma_{n,k}$ and $\Gamma_{n,k}$, it is safe to say
  \begin{subequations}
    \begin{align}
      \tilde F_k &= F_k + O(\epsilon) \\
      \tilde P_k &= P_k + O(\epsilon) \\
      \tilde H_k &= H_k + O(\epsilon^2) \\
      \tilde H_k &= O(\epsilon) \\
      H_k &= O(\epsilon)
    \end{align}
  \end{subequations}
  Therefore, (\ref{eq:close_F}), (\ref{eq:close_P}), (\ref{eq:close_H0}), (\ref{eq:close_H1}) and (\ref{eq:close_H2}) are true for $k$.

  Now that we have the results with $F_k$, $\tilde F_k$, $H_k$, $\tilde H_k$, $P_k$ and $\tilde P_k$, we can move to what are immediately following, i.e. $s_k$, $\tilde s_k$, $K_k$ and $\tilde K_k$.
  \begin{align}
    s_k =& - F_k^{-1} H_k = - F_k^{-1} O(\epsilon) = O(\epsilon)
  \end{align}
  which is true because $F_k^{-1}$ is bounded above. Similarly, we have $\tilde s_k = O(\epsilon)$. Equations (\ref{eq:close_s1}) and (\ref{eq:close_s2}) are true.

  \begin{subequations}
    \begin{align}
      \tilde s_k &= - \tilde F_k^{-1} \tilde H_k = - (F_k + O(\epsilon))^{-1} (H_k + O(\epsilon^2)) \\
                 &= - (F_k^{-1} + O(\epsilon)) (H_k + O(\epsilon^2)) \\
                 &= - F_k^{-1} H_k + F_k^{-1} O(\epsilon^2) + H_k O(\epsilon) + O(\epsilon^2) \\
                 &= s_k + O(\epsilon^2) \\
      \tilde K_k &= - \tilde F_k^{-1} \tilde P_k = - (F_k + O(\epsilon))^{-1} (P_k + O(\epsilon)) \\
                 &= - (F_k^{-1} + O(\epsilon)) (P_k + O(\epsilon)) \\
                 &= - F_k^{-1} P_k + (F_k^{-1} + P_k) O(\epsilon) + O(\epsilon^2) \\
                 &= K_k + O(\epsilon)
    \end{align}
  \end{subequations}
  Equations (\ref{eq:close_s0}) and (\ref{eq:close_K}) are true for $k$.

  Now we are equipped to get closeness/small results for $S_{n,k}$ and $\tilde S_{n,k}$.
  \begin{subequations}
    \begin{align}
      & \ \tilde S_{n,k} - S_{n,k} \nonumber \\
      =&
         \begin{bmatrix}
           1 & 0 & 0 \\
           0 & I & \tilde K_k^\top
         \end{bmatrix} \tilde \Gamma_{n, k}
                   \begin{bmatrix}
                     1 & 0 \\
                     0 & I \\
                     0 & \tilde K_k
                   \end{bmatrix} -
                         \begin{bmatrix}
                           1 & 0 & 0 \\
                           0 & I & K_k^\top
                         \end{bmatrix} \Gamma_{n, k}
                                   \begin{bmatrix}
                                     1 & 0 \\
                                     0 & I \\
                                     0 & K_k
                                   \end{bmatrix} \nonumber \\
      & +
        \begin{bmatrix}
          \tilde s_k^\top \tilde \Gamma_{n, k}^{uu} \tilde s_k + 2 \tilde s_k^\top \tilde \Gamma_{n, k}^{u1} - s_k^\top \Gamma_{n, k}^{uu} s_k - 2 s_k^\top \Gamma_{n, k}^{u1} & \mathbf{0} \\
          (\tilde \Gamma_{n, k}^{xu} + \tilde \Gamma_{n, k}^{uu} \tilde K_k) \tilde s_k - (\Gamma_{n, k}^{ux} + \Gamma_{n, k}^{uu} K_k)^\top s_k & \mathbf{0}
        \end{bmatrix} \nonumber \\
      & +
        \begin{bmatrix}
          \mathbf{0} & \tilde s_k^\top (\tilde \Gamma_{n, k}^{ux} + \tilde \Gamma_{n, k}^{uu} \tilde K_k) - s_k^\top (\Gamma_{n, k}^{ux} + \Gamma_{n, k}^{uu} K_k) \\
          \mathbf{0} & \mathbf{0}
        \end{bmatrix} \\
      =&
         \begin{bmatrix}
           \tilde \Gamma_{n,k}^{11} & \tilde \Gamma_{n,k}^{1x} + \tilde \Gamma_{n,k}^{1u} \tilde K \\
           \tilde \Gamma_{n,k}^{x1} + \tilde K^\top \tilde \Gamma_{n,k}^{u1} & \tilde \Gamma_{n,k}^{xx} + 2 \tilde \Gamma_{n,k}^{xu} \tilde K + \tilde K^\top \tilde \Gamma_{n,k}^{uu} \tilde K
         \end{bmatrix} \nonumber \\
      & -
        \begin{bmatrix}
          \Gamma_{n,k}^{11} & \Gamma_{n,k}^{1x} + \Gamma_{n,k}^{1u} K \\
          \Gamma_{n,k}^{x1} + K^\top \Gamma_{n,k}^{u1} & \Gamma_{n,k}^{xx} + 2 \Gamma_{n,k}^{xu} K + K^\top \Gamma_{n,k}^{uu} K
        \end{bmatrix} \nonumber \\
      & +
        \begin{bmatrix}
          O(\epsilon^2) & O(\epsilon^2) \\
          O(\epsilon^2) & 0
        \end{bmatrix} \\
      =&
         \begin{bmatrix}
           O(\epsilon^2) & O(\epsilon^2) \\
           O(\epsilon^2) & O(\epsilon)
         \end{bmatrix} \\
    \end{align}
  \end{subequations}
  So that \eqref{eq:close_S4}, (\ref{eq:close_S2}) and (\ref{eq:close_S3}) are true for $k$.

  \begin{subequations}
    \begin{align}
      & S_{n,k}^{1x} = M_{n,k}^{1x} + S_{n,k+1}^{1x} A_k + \Gamma_{n,k}^{1u} K_k + s_k^\top ( \Gamma_{n,k}^{ux} + \Gamma_{n,k}^{uu} K_k ) \\
      & = M_{n,k}^{1x} + \Omega_{n,k+1} A_k + A_k O(\epsilon) \nonumber \\
      & \quad + K_k O(\epsilon) +  (\Gamma_{n,k}^{ux} + \Gamma_{n,k}^{uu} K_k) O(\epsilon) \\
      & = \Omega_{n,k} + O(\epsilon)
    \end{align}
  \end{subequations}
  Therefore, (\ref{eq:close_S0}) holds and then naturally (\ref{eq:close_S1}) holds.

  We continue to prove that $\tilde D_{n,k-1}$ and $D_{n, k-1}$ are close, which is true because
  \begin{subequations}
    \begin{align}
      \tilde D_{n,k-1} =& \sum_{l=1}^{n_x} \tilde S_{n,k}^{1x^l} G_k^l \\
      =& \sum_{l=1}^{n_x} (\Omega_{n,k}^l + O(\epsilon)) G_k^l \\
      =& \sum_{l=1}^{n_x} \Omega_{n,k}^l G_k^l + O(\epsilon) \\
      =& D_{n,k-1} + O(\epsilon)
    \end{align}
  \end{subequations}
  So (\ref{eq:close_D}) is true.
  \hfill\qed
\end{pf*}

The following lemma shows that the states and actions computed in the update steps of both algorithms are close.

\begin{lemma}
  \label{lem:solClose}
  {\it
    The updates by two algorithms are small and close
    \begin{subequations}
      \label{eq:closeness_forward}
      \begin{align}
        \label{eq:DDP_approx_states}
        \delta x_{k+1}^D =& A_k \delta x_k^D + B_k \delta u_k^D + O(\epsilon^2) \\
        \label{eq:newton_approx_states}
        \delta x_{k+1}^N =& A_k \delta x_k^N + B_k \delta u_k^N + O(\epsilon^2) \\
        \label{eq:input_small_newton}
        \delta u_k^N = O(\epsilon) \\
        \label{eq:input_small_ddp}
        \delta u_k^D = O(\epsilon) \\
        \label{eq:state_small_newton}
        \delta x_k^N = O(\epsilon) \\
        \label{eq:state_small_ddp}
        \delta x_k^D = O(\epsilon) \\
        \label{eq:input_step_close}
        \delta u_k^N - \delta u_k^D =& O(\epsilon^2)\\
        \label{eq:state_step_close}
        \delta x_k^N - \delta x_k^D =& O(\epsilon^2)\\
        \label{eq:close_update_newton_DDP}
        \delta u^N - \delta u^D =& O(\epsilon^2) \\
        \label{eq:classic_newton_convergence}
        \|\bar u + \delta u^N - u^{\star}\| =& O(\epsilon^2). \\
        \label{eq:DDP_convergence}
        \|\bar u + \delta u^D - u^{\star}\| =& O(\epsilon^2).
      \end{align}
    \end{subequations}
  }
\end{lemma}

\begin{pf*}{Proof.}
Equation (\ref{eq:DDP_approx_states}) comes directly from the Taylor series expansion of (\ref{eq:dynamics}) and (\ref{eq:newton_approx_states}) from (\ref{eq:newton_dp_dynamics0}).

We prove (\ref{eq:input_small_newton}) to (\ref{eq:state_step_close}) by induction. For $k=0$, $\delta x_{0}^N = \delta x_{0}^D = 0$ and $\delta u_{0}^N = s_0,\  \delta x_{0}^D = \tilde s_0$. We know from the proof of lemma \ref{lem:matrixClose} that $s_0 = \tilde s_0 + O(\epsilon^2)$, $s_0 = O(\epsilon)$ and $\tilde s_0 = O(\epsilon)$,  so (\ref{eq:input_small_newton}) to (\ref{eq:state_step_close}) hold for $k=0$. Assume (\ref{eq:input_small_newton}) to (\ref{eq:state_step_close}) hold for $k$, then
\begin{subequations}
  \begin{align}
    \delta u^N_{k+1} &= K_{k+1} \delta x^N_k + s_{k+1} = O(\epsilon) \\
    \delta u^D_{k+1} &= \tilde K_{k+1} \delta x^N_k + \tilde s_{k+1} = O(\epsilon) \\
    \delta x^N_{k+1} =& A_k \delta x^N_k + B_k \delta u^N_k + O(\epsilon^2) = O(\epsilon) \\
    \delta x^D_{k+1} =& A_k \delta x^D_k + B_k \delta u^D_k + O(\epsilon^2) = O(\epsilon) \\
    \delta u^N_{k+1} - \delta u^D_{k+1} =& K_k \delta x^N_{k} - \tilde K_k \delta x^D_{k} + s_k - \tilde s_k \\
    =& K_k \delta x^N_k - (K_k + O(\epsilon)) (\delta x^N_k + O(\epsilon^2)) \nonumber \\
                     & + O(\epsilon^2) \\
    =& O(\epsilon) \delta x^N_k + O(\epsilon^2) \\
    =& O(\epsilon^2) \\
    \delta x^N_{k+1} - \delta x^D_{k+1} =& A_k ( \delta x^N_{k} - \delta x^D_{k} ) + B_k ( \delta u^N_{k} - \delta u^D_{k} ) + O(\epsilon^2) \nonumber \\
    =& O(\epsilon^2)
  \end{align}
\end{subequations} \\
So (\ref{eq:input_small_newton}) to (\ref{eq:state_step_close}) hold for $k+1$ and the proof by induction is done. Equation (\ref{eq:close_update_newton_DDP}) comes directly as a result. Equation (\ref{eq:classic_newton_convergence}) is classic convergence analysis for Newton's method \cite{nocedal2006numerical}. Equation (\ref{eq:DDP_convergence}) follows directly from (\ref{eq:close_update_newton_DDP}) and (\ref{eq:classic_newton_convergence}).
\hfill\qed
\end{pf*}

\balance

\end{document}